\documentclass[useAMS,usenatbib]{mn2e}
\usepackage{graphicx}
\usepackage{times}
\usepackage{mn2e-breakabs}
\usepackage{amssymb}
\usepackage{amsmath}
\begin{document}

% --- title --- %
\title[BAO and RSD of Red/Blue BOSS 3D Clustering] 
{The Clustering of Galaxies in the SDSS-III DR10 Baryon Oscillation Spectroscopic Survey: No Detectable Colour Dependence of Distance Scale or Growth Rate Measurements}

\author[A. J. Ross et al.]{\parbox{\textwidth}{
Ashley J. Ross\thanks{Email: Ashley.Ross@port.ac.uk}$^{1}$, 
Lado Samushia$^{1,2}$,
Angela Burden$^{1}$,
Will J. Percival$^{1}$,
Rita Tojeiro$^{1}$,
Marc Manera$^{1}$,
Florian Beutler$^{3}$,
J. Brinkmann$^4$
Joel R. Brownstein$^{5}$,
Aurelio Carnero$^{6,7}$
Luiz A. N. da Costa$^{6,7}$,
Daniel J. Eisenstein$^{8}$,
Hong Guo$^{5}$,
Shirley Ho$^9$,
Marcio A. G. Maia$^{6,7}$,
Francesco Montesano$^{10}$,
Demitri Muna$^{11}$,
Robert C. Nichol$^1$,
Sebasti\'an E. Nuza$^{12}$,
Ariel G. S\'anchez$^{10}$,
Donald P. Schneider$^{13,14}$,
Ramin A. Skibba$^{15}$,
Fl\'avia Sobreira$^{6,7}$,
Alina Streblyanska$^{16,17}$,
Molly E. C. Swanson$^{8}$,
Daniel Thomas$^1$,
Jeremy L. Tinker$^{18}$,
David A. Wake$^{19,20}$,
Idit Zehavi$^{21}$,
Gong-bo Zhao$^1$
}
  \vspace*{4pt} \\ 
$^{1}$Institute of Cosmology \& Gravitation, Dennis Sciama Building, University of Portsmouth, Portsmouth, PO1 3FX, UK\\
$^{2}$National Abastumani Astrophysical Observatory, Ilia State University, 2A Kazbegi Ave., GE-1060 Tbilisi, Georgia\\
$^{3}$Lawrence Berkeley National Laboratory, 1 Cyclotron Road, Berkeley, CA 94720, USA\\
$^4$Apache Point Observatory, PO Box 59, Sunspot, NM 88349-0059, USA\\
$^5$Department of Physics and Astronomy, University of Utah, UT 84112, USA\\
$^{6}$Observat\'orio Nacional, Rua Gal. Jos\'e Cristino 77, Rio de Janeiro, RJ - 20921-400, Brazil\\
$^7$Laborat\'orio Interinstitucional de e-Astronomia - LineA, Rua Gal. Jos\'e Cristino 77, Rio de Janeiro, RJ - 20921-400, Brazil\\
$^8$Harvard-Smithsonian Center for Astrophysics, 60 Garden St., Cambridge, MA 02138, USA\\
$^9$Department of Physics, Carnegie Mellon University, 5000 Forbes Avenue, Pittsburgh, PA 15213, USA\\ 
$^{10}$Max-Planck-Institut f\"ur extraterrestrische Physik, Postfach 1312, Giessenbachstra{\ss}e., 85748 Garching, Germany\\
$^{11}$Department of Astronomy, Ohio State University, Columbus, OH, 43210, USA\\
$^{12}$Leibniz-Institut f\"ur Astrophysik Potsdam (AIP), An der Sternwarte 16, D-14482 Potsdam, Germany\\
$^{13}$Department of Astronomy and Astrophysics, The Pennsylvania State University, University Park, PA 16802\\
$^{14}$Institute for Gravitation and the Cosmos, The Pennsylvania State University, University Park, PA 16802\\
$^{15}$Center for Astrophysics and Space Sciences, University of California, 9500 Gilman Drive, San Diego, CA 92093, USA\\
$^{16}$Instituto de Astrofisica de Canarias (IAC), E-38200 La Laguna, Tenerife, Spain\\
$^{17}$Universidad de La Laguna (ULL), Dept. Astrofisica, E-38206 La Laguna, Tenerife, Spain\\
$^{18}$Center for Cosmology and Particle Physics, New York University, New York, NY 10003, USA\\
$^{19}$Department of Astronomy, University of Wisconsin-Madison, 475 N. Charter Street, Madison, WI, 53706, USA\\
$^{20}$Department of Physical Sciences, The Open University, Milton Keynes MK7 6AA, UK\\
$^{21}$Department of Astronomy, Case Western Reserve University, OH 44106, USA
}

\date{Accepted for publication by MNRAS} 

\pagerange{\pageref{firstpage}--\pageref{lastpage}} \pubyear{2012}
\maketitle
\label{firstpage}

\begin{abstract}
We study the clustering of galaxies, as a function of their colour, from Data Release Ten (DR10) of the Sloan Digital Sky Survey III (SDSS-III) Baryon Oscillation Spectroscopic Survey. DR10 contains 540,505 galaxies with $0.43 < z < 0.7$; from these we select 122,967 into a ``Blue'' sample and 131,969 into a ``Red'' sample based on $k+e$ corrected (to $z=0.55$) $r-i$ colours and $i$ band magnitudes. The samples are chosen to each contain more than 100,000 galaxies, have similar redshift distributions, and maximize the difference in clustering amplitude. The Red sample has a 40\% larger bias than the Blue ($b_{Red}/b_{Blue} = 1.39\pm0.04$), implying that the Red galaxies occupy dark matter halos with an average mass that is 0.5 ${\rm log}_{10} M_{\odot}$ greater. Spherically averaged measurements of the correlation function, $\xi_0$, and the power spectrum are used to locate the position of the baryon acoustic oscillation (BAO) feature of both samples. Using $\xi_0$, we obtain distance scales, relative to the distance of our reference $\Lambda$CDM cosmology, of $1.010\pm0.027$ for the Red sample and $1.005\pm0.031$ for the Blue. After applying reconstruction, these measurements improve to $1.013\pm0.020$ for the Red sample and $1.008\pm0.026$ for the Blue. For each sample, measurements of $\xi_0$ and the second multipole moment, $\xi_2$, of the anisotropic correlation function are used to determine the rate of structure growth, parameterized by $f\sigma_8$. We find $f\sigma_{8,Red} = 0.511\pm0.083$, $f\sigma_{8,Blue} = 0.509\pm0.085$, and $f\sigma_{8,Cross} = 0.423\pm0.061$ (from the cross-correlation between the Red and Blue samples). We use the covariance between the bias and growth measurements obtained from each sample and their cross-correlation to produce an optimally-combined measurement of $f\sigma_{8,comb} = 0.443\pm0.055$. This result compares favorably to that of the full $0.43 < z < 0.7$ sample ($f\sigma_{8, full} = 0.422\pm0.051$) despite the fact that, in total, we use less than half of the number of galaxies analyzed in the full sample measurement. In no instance do we detect significant differences in distance scale or structure growth measurements obtained from the Blue and Red samples. Our results are consistent with theoretical predictions and our tests on mock samples, which predict that any colour dependent systematic uncertainty on the measured BAO position is less than 0.5 per cent.
\end{abstract}


\section{Introduction}

In the last decade, wide-field galaxy redshift surveys such as the Two Degree Field Galaxy Redshift Survey (2dFGRS; \citealt{2df}), the Sloan Digital Sky Survey (SDSS; \citealt{SDSS}), and the WiggleZ Dark Energy Survey \citep{wiggz} have provided a wealth of information for cosmological analyses (e.g., \citealt{Teg04,2dfcos,Eis05,Per09BAO,ReidDR7,BlakeKazinBAO,MontesanoPk,alph,Reid12RSD,Sanchez12}). In particular, measurements of the baryon acoustic oscillation (BAO) scale and of the redshift-space distortion (RSD) signal enable measurements of Dark Energy (see, e.g., \citealt{WeinbergDE} for a review) and allow tests of General Relativity (see, e.g., \citealt{JZ08,SP09}). 

All studies that wish to use the distribution of galaxies to measure cosmological parameters must account for any uncertainty in the manner with which galaxies trace the underlying matter distribution. Galaxies are observed over many orders of magnitude in luminosity and have a bimodal colour distribution (see, e.g. \citealt{BlantonEtAl03,Baldry04,Bell04}). The clustering of galaxies as a function of luminosity and colour has been extensively studied (see, e.g., \citealt{W98,N02,Ma03,Z05,Cr06,Li,R07,M08,swanson,cresswell,SS09,R09,T10,Z11,Christ12,Guo13,Hartley13,Skibba13}). Generally, it has been found that the clustering strength, parameterized as the `bias', increases in the direction of greater luminosity and redder colour, and that colour is more predictive of the large-scale environment than other characteristics such as morphology (\citealt{Ball08,Skibba09}).

The observed colour and luminosity dependence of galaxy clustering is consistent with a model in which the more luminous and red galaxies occupy dark matter halos of the greater mass. In the widely accepted model of galaxy evolution, galaxies form within the gravitational potential wells of host dark matter halos \citep{WhiteRees1978}. The large-scale clustering of the halos, and thus the galaxies that reside in them, is directly linked to the dark matter halo mass \citep{BBKS,cole89}. This model, that the clustering of galaxies is determined solely by the mass of the halos they occupy, has provided an excellent description of the locally observed galaxy distribution (see, e.g., \citealt{N02,Z11}), the distribution in the distant Universe (see, e.g., \citealt{Coil06,Mc07}) and indeed the distribution of subsets of the galaxy sample we use in this study (\citealt{White11BEDR,Nuza12}). Further, it has been found that the observed distribution of the total galaxy population and its subsets split by colour can self-consistently be described by a model that depends only on halo mass (see, e.g., \citealt{Z05,Tinker08,R09,SS09,T10,Z11}).

The results described above suggest that the large-scale clustering of galaxies depends only on the mass of the halos they occupy. Under this assumption, in order to test methods of measuring cosmological parameters from galaxy clustering, one only requires simulations that are able produce clustering statistics as a function of halo mass. Using a combination of results from perturbation theory and simulation, \cite{Eis07,Angulo08,Pad09,Mehta11,SZ12,McCullagh13} suggest a bias-dependent shift in the BAO position that is less than 0.5 per cent. \cite{RW11} studied the RSD signal as a function of halo mass and produced a model accurate to within 2\% for tracers with bias $\sim$2 and scales greater than 40 $h^{-1}$ Mpc.

In our study, we measure the clustering of galaxies from data release ten (DR10; \citealt{DR10}) of the SDSS-III Baryon Oscillation Spectroscopic Survey (BOSS; \citealt{Daw12bosso}) when divided by colour. Dividing the sample by colour provides a straightforward way to separate the data into two samples with different bias (and implied halo mass). We expect that any physical properties (e.g., collapse time, shape) of the halo other than its mass that are required to understand the large-scale clustering of galaxies will correlate with the differences in the intra-galaxy physical processes, clearly observed via the difference in the colour of the galaxies' stellar populations. We are therefore undertaking the most simple binary test that probes whether BAO and RSD measurements are robust to the galaxy sample that is chosen to make the measurement.

We use data from the DR10 BOSS `CMASS' sample. Measurements of the clustering of BOSS CMASS galaxies have been shown to be robust to many potential systematic concerns \citep{Ross12} and the measurements using data from the SDSS-III data release nine \citep{DR9} have already been used in many cosmological analyses \citep{alph,Reid12RSD,Sanchez12,Tojeiro12cos,Zhao12neut,Planck16,Anderson13,Chuang13,Kazin13,Ross_fnl,Scoccola13,SamDR9,Sanchez13}.

We analyze the distributions of our galaxy samples in both configuration space, via the multipoles of the redshift-space correlation function $\xi_{0,2}(s)$, and Fourier space, via the spherically-averaged power spectrum, $P(k)$. We describe the real and simulated data analyzed in this investigation in Section \ref{sec:data}. We describe the modeling we use in Section \ref{sec:mod}. We display the clustering measurements and the comparison between results obtained from the simulated and real data sets in Section \ref{sec:meas}. We present our BAO measurements in Section \ref{sec:bao} and our RSD measurements in Section \ref{sec:RSD}. We present our concluding remarks in Section \ref{sec:con}. We assume a flat $\Lambda$CDM cosmology with $\Omega_m=0.274$, $\Omega_bh^2 = 0.024$, $h=0.7$, $\sigma_8 = 0.8$ (same as used in, e.g. \citealt{alph}) unless otherwise noted.

\section{Data}
\label{sec:data}
The SDSS-III BOSS \citep{Eisenstein11,Daw12bosso} obtains targets using SDSS imaging data. In combination, the SDSS-I, SDSS-II, and SDSS-III surveys obtained wide-field CCD photometry (\citealt{C,Gunn06}) in five passbands ($u,g,r,i,z$; \citealt{F}), amassing a total footprint of 14,555 deg$^2$, internally calibrated using the `uber-calibration' process described in \cite{Pad08}, and with a 50\% completeness limit of point sources at $r = 22.5$ (\citealt{DR8}). From this imaging data, BOSS has targeted 1.5 million massive galaxies, 150,000 quasars, and over 75,000 ancillary targets for spectroscopic observation over an area of 10,000 deg$^2$ \citep{Daw12bosso}. BOSS observations began in fall 2009, and the last spectra of targeted galaxies will be acquired in mid-2014. The BOSS spectrographs (R = 1300-3000) are fed by 1000 optical fibres in a single pointing, each with a 2$^{\prime\prime}$ aperture \citep{Smee12}. Each observation is performed in a series of 15-minute exposures and integrated until a fiducial minimum signal-to-noise ratio, chosen to ensure a high redshift success rate, is reached. Redshifts are determined as described in \cite{Bolton12}.

We use data from the SDSS-III DR10 BOSS `CMASS' sample of galaxies, as defined by \cite{Eisenstein11}. The CMASS sample is designed to be approximately stellar mass limited above $z = 0.45$. Such galaxies are selected from the SDSS DR8 \citep{DR8} imaging via
\begin{eqnarray}
 17.5 < i_{cmod}  < 19.9\\
r_{mod} - i_{mod}  < 2 \\
d_{\perp} > 0.55 \label{eq:hcut}\\
i_{fib2} < 21.5\\
i_{cmod}  < 19.86 + 1.6(d_{\perp} - 0.8) \label{eq:slide}
\end{eqnarray}
where all magnitudes are corrected for Galactic extinction (via the \citealt{SFD} dust maps), $i_{fib2}$ is the $i$-band magnitude within a $2^{\prime \prime}$ aperture, the subscript $_{mod}$ denotes `model' magnitudes \citep{EDR}, the subscript $_{cmod}$ denotes `cmodel' magnitudes \citep{DR2}, and 
\begin{equation}
d_{\perp} = r_{mod} - i_{mod} - (g_{mod} - r_{mod})/8.0.
\label{eq:dp}
\end{equation}
This selection yields a sample with a median redshift $z = 0.57$ and a stellar mass that peaks at ${\rm log}_{10}(M/M_{\odot}) = 11.3$ \citep{Maraston12}. As the sample contains galaxies with greatest stellar mass, the majority of the sample consists of galaxies that form the red sequence. However, roughly one quarter of the galaxies would be considered `blue' by traditional SDSS (rest-frame) color cuts (see, e.g., \citealt{Strateva02}). Indeed, \cite{Masters11} find that 26\% of CMASS galaxies have a late-type (i.e., spiral disc) morphology. Like all CMASS galaxies, these blue galaxies are at the extreme end of the stellar mass function, and thus are significantly more biased than the emission line galaxies observed by WiggleZ at similar redshifts \citep{Blake10}. See \cite{Tojeiro12kcorr} for a detailed description of the CMASS population of galaxies.

We use the DR10 CMASS sample and, treat it in the same way as in \cite{alph}, with the exception that the treatment of systematic weights has been improved, as described in Section \ref{sec:wsys}. The sample has 540,505 galaxies with $0.43 < z < 0.7$ spread over an effective area of 6516 deg$^2$, 5105 deg$^2$ of which is in the North Galactic Cap. A detailed description can be found in \cite{Aardwolf}. 

\begin{figure}
\includegraphics[width=84mm]{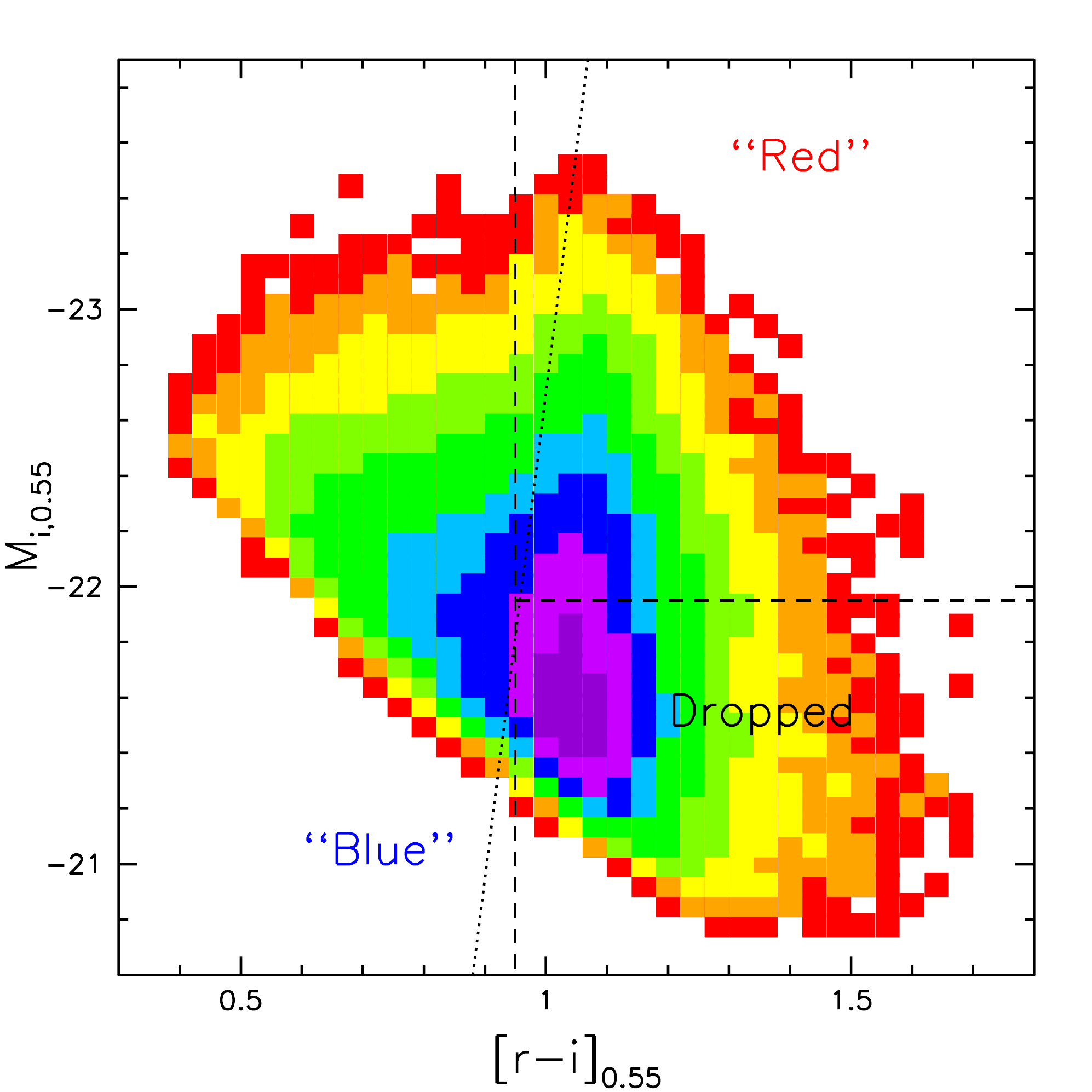}
  \caption{The density of CMASS galaxies in $i$-band absolute magnitude, $r-i$ colour space, with both values $k$-corrected to $z=0.55$. Dashed lines display the cuts we apply to define our ``Red'', containing 131,969 galaxies, and ``Blue'', containing 122,967 galaxies, samples. The 285,569 galaxies occupying the lower-right box (labeled ``Dropped'') are not included in our analysis. The dotted diagonal line displays the cut applied by Guo et al. (2013) to separate ``blue'' and ``green'' BOSS galaxies with $0.55 < z < 0.6$ from the ``red'' and ``reddest''.}
  \label{fig:abrisel}
\end{figure}

\begin{figure}
\includegraphics[width=84mm]{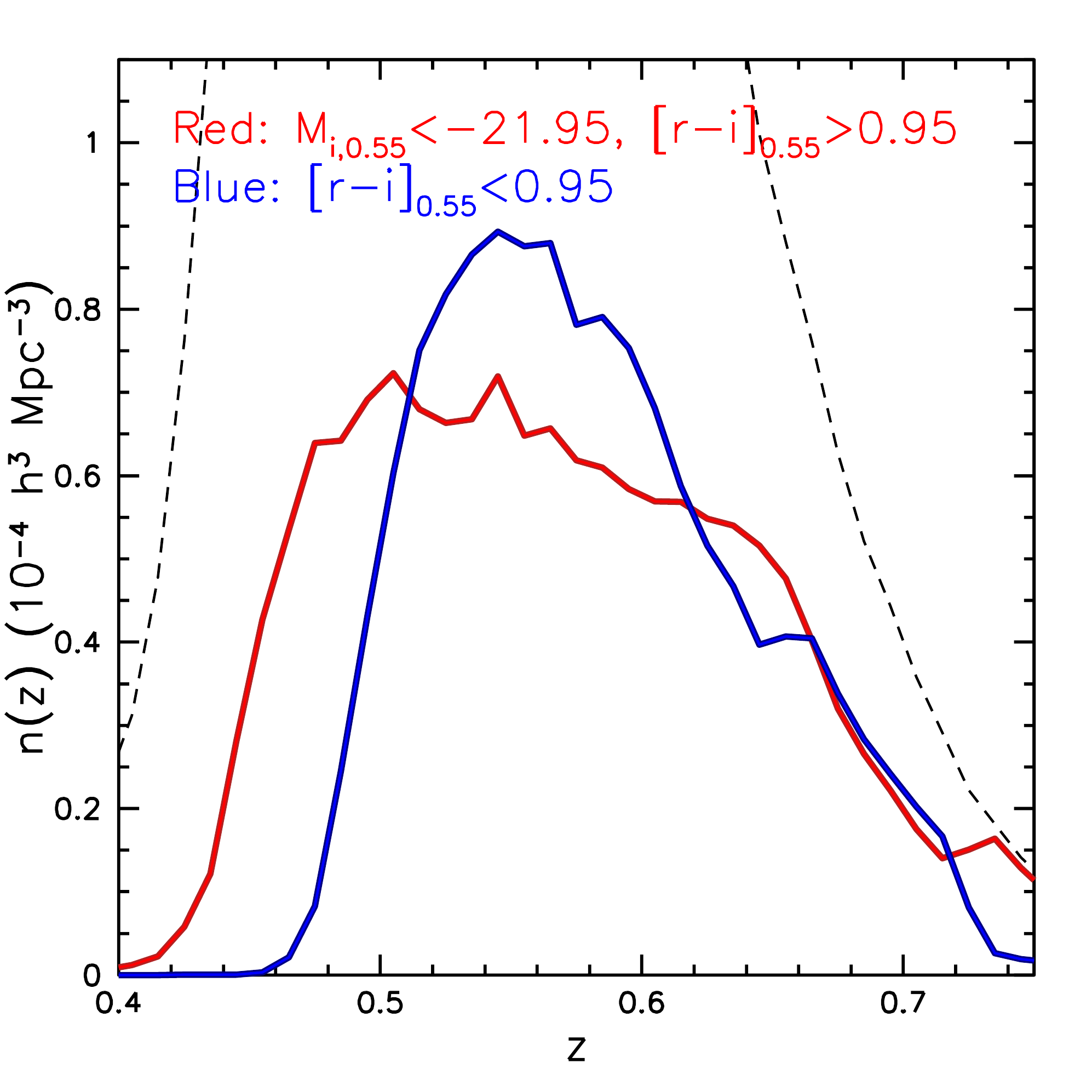}
  \caption{The $n(z)$ of our Red and Blue samples. The dashed black line shows the $n(z)$ of the full CMASS sample. The number of Red and Blue galaxies is each approximately one quarter the number of the full sample, but each make up half of the full sample for $z > 0.65$.}
  \label{fig:nz}
\end{figure}

\subsection{Dividing the BOSS Galaxies by Colour}
\label{sec:coldef}
In order to compare CMASS galaxies at different redshifts, we apply corrections to the measured magnitudes in order to account for the redshifting of spectral energy distribution ($k$-corrections) and the evolution of stellar populations within galaxies ($e$-corrections). We use the $k+e$ corrections of \cite{Tojeiro12kcorr} to obtain $z=0.55$ galaxy colours and magnitudes. These $k+e$ corrections were computed based on the average spectral evolution of SDSS-I/II Luminous Red Galaxies (LRGs, $\langle z\rangle = 0.35$), and were shown to describe the colours and evolution of LRGs and CMASS galaxies on average. They are applied using a single template, as a function of redshift. All corrections are computed to $z_c=0.55$ (close to the median redshift of CMASS galaxies) and refer to filters shifted to the same redshift. The rest-frame $k+e$ correction in any given band for a galaxy at $z_c=0.55$ is thus independent of the modelling and equal to $-2.5 \log_{10} [1/(1+z_c)]$. The $k+e$ corrections we use were derived based on the stellar population synthesis models of \cite{CGW09} and \cite{CG10}. \cite{Tojeiro12kcorr} also derived $k+e$ corrections based on the stellar population models of \cite{Maraston11SP}. The smaller redshift span of the galaxies in this paper combined with our choice of shifted filters and $z_c$ reduces the dependence of the $k+e$ corrections on the underlying stellar population models when compared to \cite{Tojeiro12kcorr}, and we therefore expect our conclusions to be robust to the choice of stellar population modeling. The $k+e$ corrections we employ have previously been applied to create the subsets of the CMASS galaxy sample used for the clustering studies of both \cite{Tojeiro12cos} and \cite{Guo13}.

Our aim is to balance the following three concerns when defining our samples: \begin{enumerate} 
\item Split the CMASS sample by colour to produce a bluer sample that includes the maximum number of galaxies (in order to minimize shot-noise) that are not clearly members of the red sequence
\item produce two samples with similar $n(z)$, so that that the cosmic variance of the underlying structure is the same
\item maximize the difference in clustering amplitude between the two samples. 
\end{enumerate} 
Balancing these concerns leads us to define a ``Blue'' sample as CMASS galaxies with $[r-i]_{0.55} < 0.95$ and a ``Red'' sample as CMASS galaxies with $[r-i]_{0.55} > 0.95$ and $M_{i,0.55} < -21.95$. The colour magnitude selection is displayed in Fig. \ref{fig:abrisel}. The absolute magnitude cut removes lower-luminosity galaxies that are predominantly at lower redshift. This cut thus improves the match between the $n(z)$ of the two samples and increases the difference in clustering amplitude. 

Our Blue sample has 122,967 galaxies and the Red 131,969, combining to make up 47 per cent of the total CMASS sample. The Blue sample is similar to the one obtained by applying the \cite{Masters11} observed-frame $g-i < 2.35$ cut, but we find applying the cut based on $[r-i]_{0.55}$ colour yields a larger separation in clustering amplitude and a better overlap in redshift. Our colour cut is close to the one of \cite{Guo13}, who used $[r-i]_{0.55}$ colour cuts (and the same $k+e$ corrections), that separate ``blue'' and ``green'' galaxies from the ``red'' and ``reddest'' samples. The cut applied in \cite{Guo13} for galaxies with $0.55 < z < 0.6$ is displayed with a dotted line in Fig. \ref{fig:abrisel}. We find applying such cut yields similar clustering results to the one we have chosen, but the $n(z)$ disagree slightly more and the difference in clustering amplitude is slightly smaller. This is due to the fact that we have not created volume-limited samples and thus the main effect of such a cut on our samples is to shift high luminosity galaxies from the ``Red'' sample to the ``Blue'' one.

The $n(z)$ for each sample are shown in Fig \ref{fig:nz}; they appear similar, and we test this by quantifying the effective redshift, $z^{eff}$, of each sample and the overlap, $o_{\rm Blue,Red}$, of the $n(z)$. We find the pair-weighted $z^{eff}$ as a function of the separation $s$
\begin{equation}
z^{eff}(s) = \frac{DD_z(s)}{DD(s)},
\label{eq:zeff}
\end{equation}
where the $DD$ pair-counts are weighted in the same manner as for calculating $\xi(s)$, as described in Section \ref{sec:wsys}. For $DD_z$, each pair-count is weighted by $(z_1+z_2)/2$. In the range $20 < s < 200$, we find $z^{eff}$ is nearly constant for both samples and $z^{eff}_{\rm Blue} = 0.585$ and $z^{eff}_{\rm Red} = 0.570$.
We define the overlap as 
\begin{equation}
o_{1,2} = 1-\frac{\int dV \left(n_{1}(z)-n_{2}(z)\right)^2}{\int dV n^2_{1}(z)+\int dVn^2_{2}(z)}
\end{equation}
 and find $o_{\rm Blue,Red} = 0.93$. Compared to the full CMASS sample, we find $o_{\rm full,Red} = 0.40$ and $o_{\rm full,Blue} = 0.37$. Large overlap is ideal for testing differences between the samples, as the cosmic variance of the underlying structure should be same, i.e., the differences in our measurements will be due mainly to shot-noise. In addition to improving our ability to test for systematic differences in results obtained with each sample (as compared to samples occupying independent volumes), this will also allow us to test if the RSD measurements can be improved by having two tracers, using methods similar to those described in \cite{McDSel09}.

\subsection{Systematic Weights}
\label{sec:wsys}
As in \cite{alph}, we correct for systematic trends in the observed number density of CMASS galaxies and our ability to measure redshifts using a series of weights. As in \cite{imsys} and \cite{Ross12}, we have performed tests against the potential systematics of stellar density, Galactic extinction, seeing, airmass, and sky background that may affect the DR8 \citep{DR8} imaging data used to select the (full) DR10 CMASS sample. We correct for the systematic relationship between the number density of galaxies as a function $i_{fib2}$ and stellar density using weights, $w_{\rm star}$, defined in the same manner as \cite{alph}
\begin{equation}
w_{\rm star} (n_{\rm s}, i_{\rm fib2}) = A(i_{\rm fib2}) + B(i_{\rm fib2}) n_s.
\label{eq:wstar}
\end{equation}
The coefficients $A(i_{\rm fib2})$ and $B(i_{\rm fib2})$ are given in \cite{Aardwolf} and are empirically determined using the full DR10 CMASS sample. We follow \cite{Aardwolf} and include an additional weight, $w_{\rm see}$, to correct a systematic relationship observed between the number density of CMASS galaxies and the seeing, $S$:
\begin{equation}
w_{\rm see} (S) = A\left[1-{\rm erf}\left(\frac{S-B}{\sigma}\right)\right]^{-1},
\label{eq:wsee}
\end{equation}
where the values $A=1.03$, $B=2.09$, and $\sigma=0.731$ were empirically determined using the full DR10 CMASS sample. Further details can be found in \cite{Aardwolf}. As in \cite{alph}, we correct for fibre-collided close pairs and redshift failures by increasing the weight of the nearest CMASS target by 1, we denote these weights $w_{cp}$ and $w_{zf}$, respectively. 

\begin{figure}
\includegraphics[width=84mm]{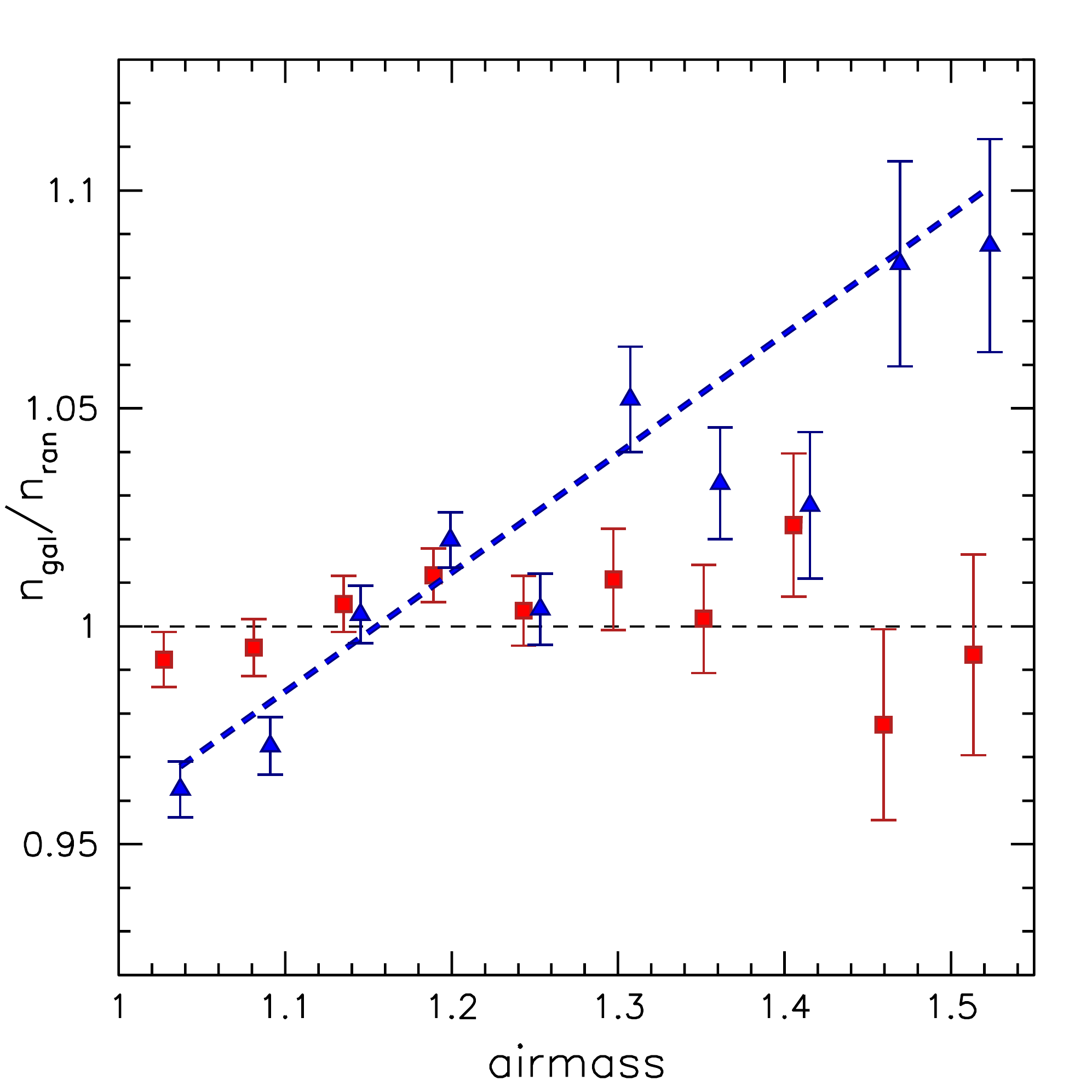}
  \caption{The relationship between the observed number density and the airmass of the imaging data used to select our Red and Blue samples. The dashed blue line displays the best-fit linear relationship for the Blue sample. The inverse of this fit is used to apply a weight to correct for the systematic effect this relationship has on the measured clustering of the Blue sample.}
  \label{fig:ngvair}
\end{figure}

For the Red and Blue samples, we have individually repeated the tests against potential systematics. We find that the systematic weights calibrated using the full sample, effectively remove the relationships with stellar density and seeing in our Red and Blue samples. However, the number density of the Blue sample has a substantial correlation with the airmass at the time the imaging data was observed; the effect is displayed in Fig. \ref{fig:ngvair}. One expects such behaviour due to the fact that magnitude errors increase at higher airmass, and thus there will be greater scatter (due to Eddington bias) across any colour cut at greater airmass. Further, we may expect airmass to have a greater effect on bluer objects, as the airmass is related to atmospheric extinction and thus greater at shorter wavelengths.

Our tests suggest that the systematic relationship with airmass is a consequence of the colour cut and therefore does not affect the density field of the full CMASS sample. For example, we do not find that the distribution of CMASS galaxies at the high redshift end of the sample has a systematic dependence with airmass, even though the Blue galaxies make up approximately half of the full sample for $z>0.65$ (as can be seen by comparing the red and blue curves in Fig. \ref{fig:nz} to the dashed curve).

\begin{figure}
\includegraphics[width=84mm]{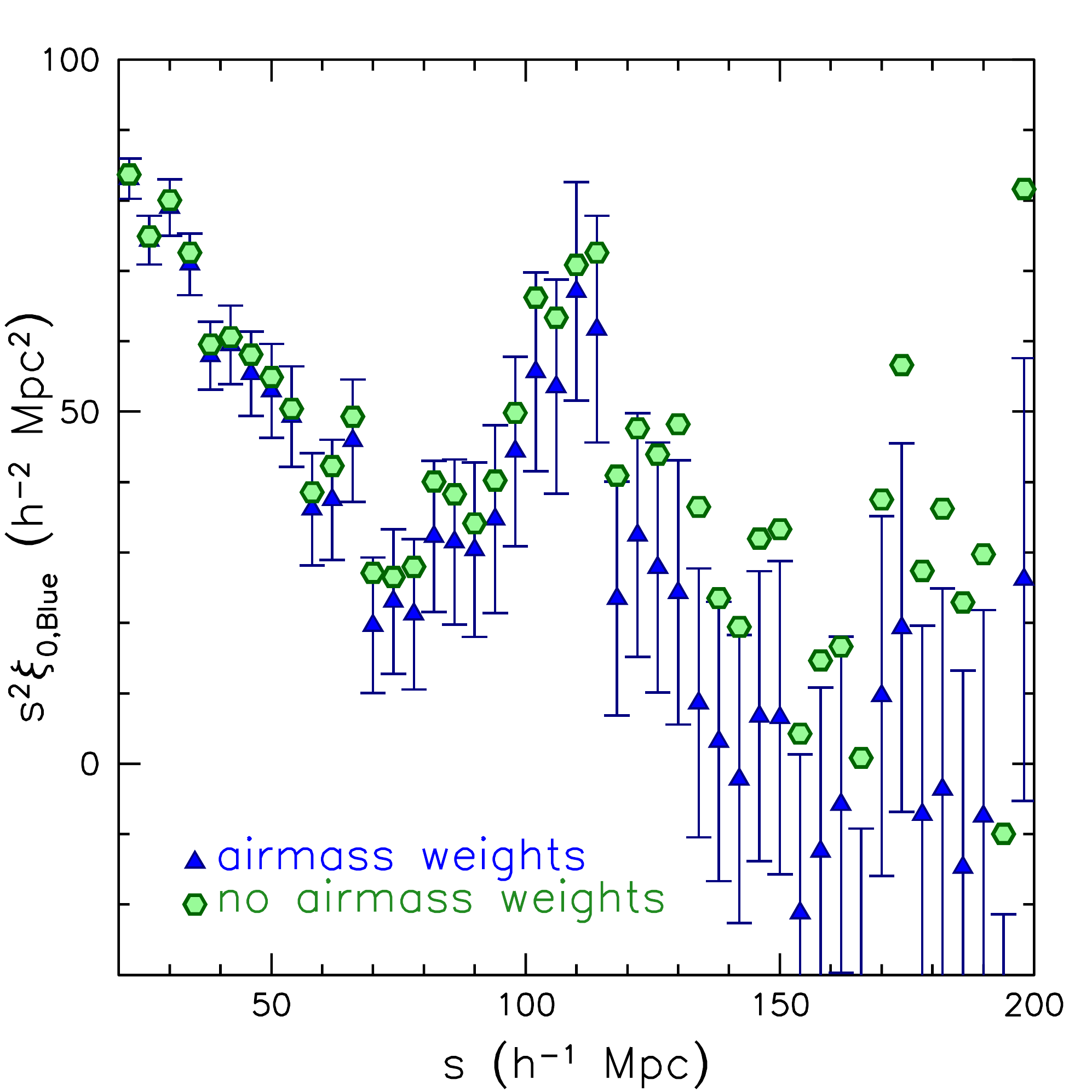}
  \caption{The measured $\xi_0$ for the Blue sample when applying weights that correct for the relationship with airmass (blue) and when these weights are not applied (green). Applying the weights changes the measured clustering by more than 1$\sigma$ for $s > 120 h^{-1}$Mpc, demonstrating their importance. }
  \label{fig:bluewair}
\end{figure}

The best-fit linear relationship between the number density of the Blue sample and the airmass, $am$, is, displayed using a dashed curve in Fig. \ref{fig:ngvair}. The inverse of this fit is used to define a weight, $w_{\rm air}$, given by
\begin{equation}
w_{\rm air} = \frac{1}{0.687 + 0.273am}.
\label{eq:wair}
\end{equation}
 Fig. \ref{fig:bluewair} displays the effect that applying $w_{\rm air}$ has to the Blue correlation function monopole. The relative effect increases towards larger scales and is greater than 1$\sigma$ for $s > 120 h^{-1}$Mpc. The total systematic weight, $w_{\rm sys}$, we apply to each galaxy is
 \begin{equation}
 w_{\rm sys} = (w_{cp}+w_{zf}-1)w_{\rm star}w_{\rm see}w_{\rm air},
 \label{eq:wsys}
 \end{equation}
 where $w_{\rm air} = 1$ for the Red sample.

\subsection{Creating Mock Galaxy Samples and Covariance Estimates}

We use 600 mock galaxy catalogs created using the PThalos methodology described in \cite{DR9mocks} to simulate the BOSS DR10 CMASS sample. Each mock catalog is masked to the same area and down-sampled to have the same mean $n(z)$ as the DR10 CMASS sample. In addition, each mask sector of each mock realization is down-sampled based on the fraction of fiber collisions, redshift failures, and completeness of the observed CMASS sample. Further details can be found in \cite{DR9mocks} and \cite{Aardwolf}.

Each mock catalog simulates the full CMASS data set. Due to the mass resolution of the matter fields, approximately 25\% of the galaxies in each (full CMASS) mock catalog are assigned to the positions of field matter particles (see \citealt{DR9mocks}). In what follows, we treat these galaxies as residing in halos with $M_{{\rm halo}} <10^{12.3} h^{-1} M_{\odot}$. In order to divide each catalog into ``Red'' and ``Blue'' mock galaxy samples we first measure $\xi_0(s)$ of ten mock samples with mass thresholds in the ranges $M_{{\rm halo}} <10^{12.3} h^{-1} M_{\odot}$ to $M_{{\rm halo}} <10^{13} h^{-1} M_{\odot}$ and $M_{{\rm halo}} >10^{13.4} h^{-1} M_{\odot}$ to $M_{{\rm halo}} >10^{14} h^{-1} M_{\odot}$ using steps of 0.05${\rm log}_{10}(M_{\odot})$ in halo mass. Using the variance of the $\xi_0(s)$ measurements as a diagonal covariance matrix, we then find the mass thresholds that yield the best-fit when comparing the mock $\xi_0(s)$ to the measured $\xi_0(s)$ of the Red and Blue samples, fitting in the range $30 < s < 100 h^{-1}$Mpc. We find $M_{{\rm halo}} < 10^{13.6} h^{-1} M_{\odot}$ for the Blue sample and $M_{{\rm halo}} > 10^{12.7} h^{-1} M_{\odot}$ for the Red sample. This implies mock galaxies residing in halos with $10^{12.7} < M_{{\rm halo}} < 10^{13.6} h^{-1} M_{\odot}$ must be split between the Red and Blue samples. We then subsample each mass-threshold sample to match the observed $n(z)$ of the Red and Blue samples in a manner that ensures that each mock galaxy is only assigned to at most one (i.e., not both) of the samples. This approach results in Blue samples that have 30\% field matter particles and Red samples that are only composed of mock galaxies within halos. In Section \ref{sec:RMcom}, we show that the mean clustering of the 600 mock samples remains well-matched to the data when the full covariance matrix is taken into account. 

\begin{figure}
\includegraphics[width=84mm]{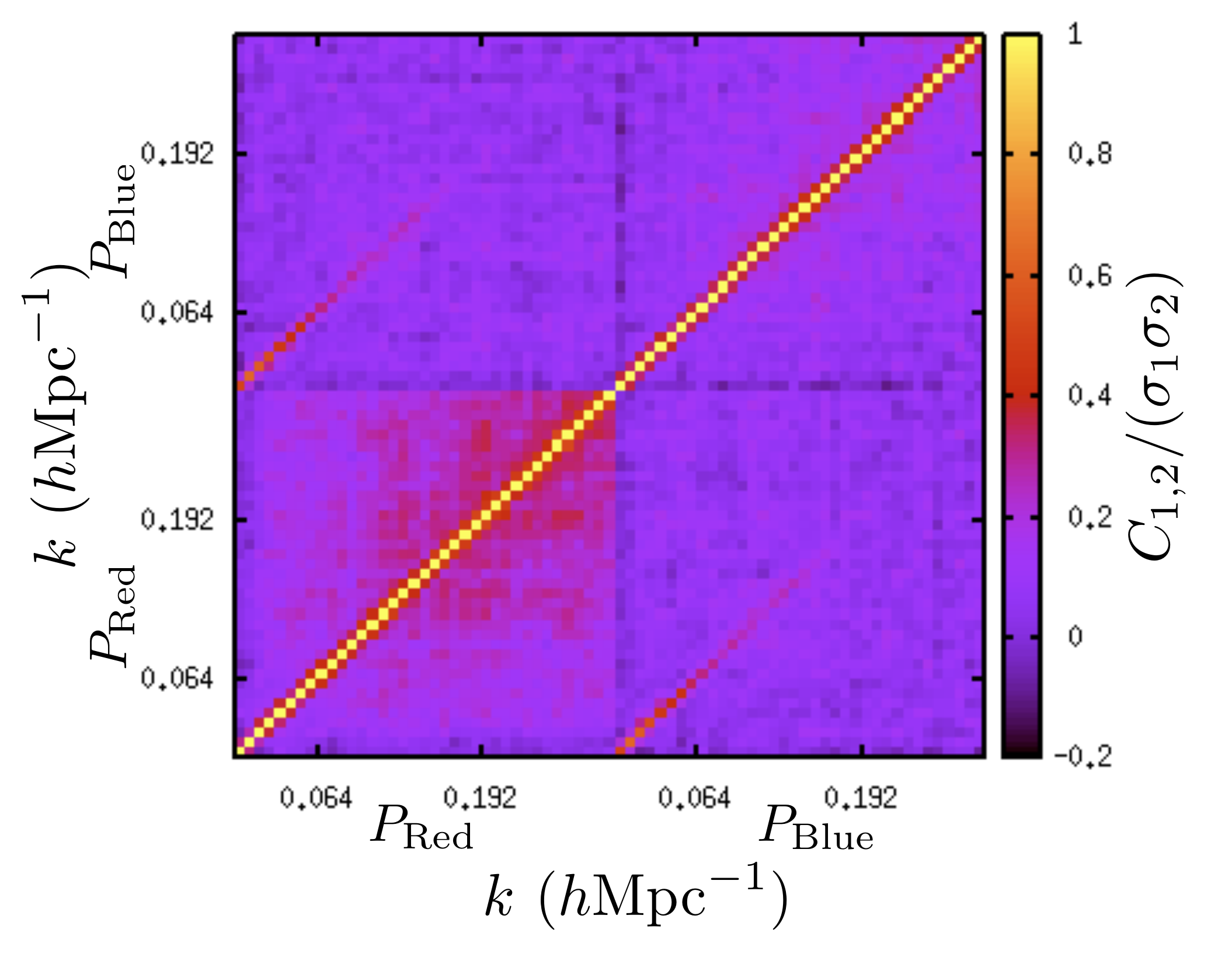}
  \caption{ The normalized covariance of the spherically averaged power spectrum, $P$, for the Red and Blue samples, determined using 600 mock realizations of each sample. The covariance matrices are close to diagonal and the covariance between the Red and Blue $P(k)$ is significantly scale dependent.}
  \label{fig:covallpk}
\end{figure}

\begin{figure}
\includegraphics[width=84mm]{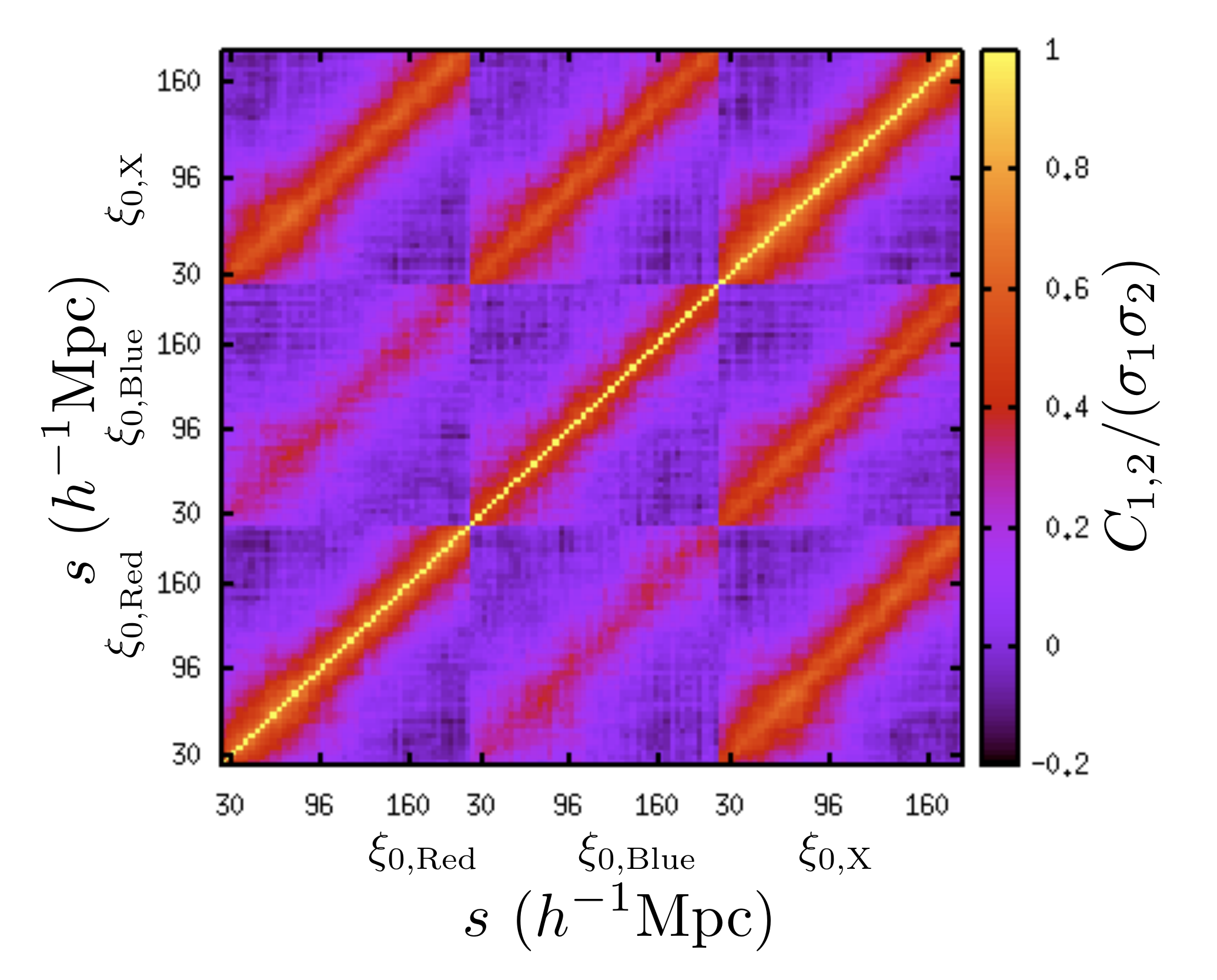}
  \caption{ The normalized covariance of the monopole of redshift-space correlation function, $\xi_0$, for the Red and Blue samples and their cross-correlation, determined using 600 mock realizations of each sample. Significant covariance is present between off-diagonal elements, as expected for $\xi_0$.}
  \label{fig:covall}
\end{figure}

For each of the 600 mock Red and Blue catalogs, we calculate the auto- and
cross- clustering statistics $\xi_{0.2}$ and $P(k)$, as described in Section \ref{sec:meas}. The estimated covariance
$\tilde{C}$ between statistic $X$ in measurement bin $i$ and statistic $Y$ in
measurement bin $j$ is then
\begin{equation}
\tilde{C}(X_i,Y_j) = \frac{1}{599}\sum_{m=1}^{600}(X_{i,m}-\bar{X}_i)(Y_{j,m}-\bar{Y}_j).
\label{eq:cov}
\end{equation}  
Fig \ref{fig:covallpk} displays the normalized covariance of and between $P(k)$ for the Red and Blue samples Fig. \ref{fig:covall} shows the same information for $\xi(s)$, with the additional inclusion of information from the cross-correlation between the Red and Blue samples. One see that there is significantly more off-diagonal covariance in the $\xi(s)$ measurements than for $P(k)$, as expected. The covariance between the Red and Blue measurements is more scale dependent for $P(k)$ than for $\xi(s)$and their cross-correlation and we discuss this further in Section \ref{sec:bao}.

To obtain an unbiased estimate of the inverse covariance matrix $\textbf{{\sf C}}^{-1}$ we rescale
the inverse of our covariance matrix by a factor that depends on the number of
mocks and measurement bins (see e.g., \citealt{Hartlap07}) 
\begin{equation} \textbf{\sf C}^{-1} =
  \frac{N_{mocks}-N_{bins}-2}{N_{mocks}-1}~\tilde{\textbf{\sf C}}^{-1}.  \label{eq:cinv}
\end{equation}
$N_{mock}$ is 600 in all cases, but $N_{bins}$ will change depending on the
specific test we perform. We determine $\chi^2$ statistics in the standard manner,
i.e.,  
\begin{equation}
\chi^2 = (\textbf{\it X}-\textbf{\it X}_{mod}) \textbf{\sf C}_{X}^{-1} (\textbf{\it X}-\textbf{\it X}_{mod})^{T},
\end{equation}
where the data/model vector $\textbf{{\it X}}$ can contain any combination of clustering measurements. Likelihood distributions, ${\cal L}$, are determined by assuming ${\cal L}(\textbf{\it X}) \propto e^{-\chi^2(X)/2}$. 

Building from the results of \cite{DS12}, \cite{PerCov} show that there are additional factors one must apply to uncertainties determined using a covariance matrix that is constructed from a finite number of realizations and to standard deviations determined from those realizations. Defining
\begin{equation}
A = \frac{1}{(N_{mocks}-N_{bins}-1)(N_{mocks}-N_{bins}-4)},
\end{equation}
and
\begin{equation}
B = \frac{N_{mocks}-N_{bins}-2}{A},
\end{equation}
the variance estimated from the likelihood distribution should be multiplied by
\begin{equation}
m_{\sigma} = \frac{1+B(N_{bins}-N_p)}{1+2A+B(N_p+1)},
\end{equation}
and the sample variance should be multiplied by
\begin{equation}
m_{v} = m_{\sigma}\frac{N_{mocks}-1}{N_{mocks}-N_{bin}-2}.
\end{equation}
We apply these factors, where appropriate, to all values we quote.

\section{Modelling and Fitting}
\label{sec:mod}

\subsection{Spherically-averaged BAO}
\label{sec:baomod}
The modelling we employ to extract the BAO position from $\xi(s)$ and $P(k)$
measurements is based on the techniques applied by \cite{Xu12mon} and \cite{alph}. We have made some slight modifications that make the
approach to each observable more consistent. For each observable, we extract a dilation
factor $\alpha$ from a template that includes the BAO feature, relative to a
smooth shape that has considerable freedom. Assuming spherical symmetry, the
measurement of $\alpha$ can be related to physical distances via
\begin{equation} 
  \alpha = (D_V(z)/r_s)/(D_V(z)/r_s)_{fid}, \label{eq:alphsph}
\end{equation}
where
\begin{equation}
D_V(z) \equiv \left((1+z)^2D^2_A(z)\frac{cz}{H(z)}\right)^{1/3}
\label{eq:Dv}
\end{equation}
and $r_s$ is the sound horizon at the baryon drag epoch, which can be accurately calculated using, e.g., the software package {\sc CAMB}\footnote{camb.info} (see, e.g., \citealt{Planck16} and references therein), and $D_A(z)$ is the angular diameter distance. 

We obtain the linear power spectrum, $P_{\rm lin}(k)$, using {\sc CAMB}. We obtain the power spectrum with no BAO feature, $P_{\rm smooth}(k)$, using the fitting formulae of \cite{EH98}. As in, e.g., \cite{Xu12mon,alph}, we then define
\begin{equation}
P_{\rm BAO}(k) = \left[P_{\rm lin}(k)-P_{\rm smooth}(k)\right]{\rm e}^{-k^2\Sigma^2_{nl}/2},
\end{equation}
where $\Sigma_{nl}$ accounts for the smearing of the BAO feature due to non-linear effects. We transform $P_{\rm smooth}$ and $P_{\rm BAO}$ to obtain $\xi_{\rm smooth}$ and $\xi_{\rm BAO}$ via
\begin{equation}
\xi(s) = \frac{1}{2\pi^2}\int dk k^2 \frac{{\rm sin}(ks)}{ks} P(k){\rm e}^{-k^2},
\end{equation}
where the e$^{-k^2}$ damps the integrand at large $k$, improving convergence of the integral without decreasing accuracy at scales relevant to the BAO feature \citep{Xu12mon}.

We model both $\xi(s)$ and $P(k)$ with the use of the smooth component plus a three parameter polynomial. For $P(k)$, we use
\begin{equation}
P_{\rm mod}(k) = P_{\rm NoBAO}(k)\frac{\left(1+P_{\rm BAO}(k/\alpha,\Sigma_{nl})\right)}{P_{\rm smooth}(k/\alpha)}
\end{equation}
where
\begin{equation}
P_{\rm NoBAO}(k) = B^2_pP_{\rm smooth}(k) +\frac{C_0}{k}+\frac{C_1}{k^2}+\frac{C_2}{k^3}.
\end{equation}
The $P_{mod}$ is then convolved with the window function and compared with the measured $P(k)$. The three-term polynomial is the Fourier equivalent of the three-term polynomial applied by \cite{Xu12mon} and \cite{alph} to fit $\xi(s)$. We find no significant changes when an extra constant term is included in the model. The $P(k)$ model is divergent at low $k$, but we find it provides a good description of our measurements over the range in scales that we fit to obtain the BAO scale ($0.02 < k < 0.3 h$Mpc$^{-1}$).

The model for the correlation function is expressed as
\begin{equation}
\xi_{\rm mod}(s) = \xi_{\rm NoBAO}(\alpha s)+B_{B}\xi_{\rm BAO}(\alpha s)
\label{eq:xialph}
\end{equation}
where 
\begin{equation}
\xi_{\rm NoBAO}(s) = B^2_{\xi}\xi_{\rm smooth}(\alpha s)+A_0+\frac{A_1}{\alpha s}+\frac{A_2}{(\alpha s)^2}
\label{eq:xialph}
\end{equation}
and we use $B_{B}$ to keep the relative height of the BAO peak constant via
\begin{equation}
B_{B} = \frac{\xi_{\rm NoBAO}(s_B)}{\xi_{\rm smooth}(s_B)}.
\end{equation}
We choose $s_{B}$ to be $\alpha50 h^{-1}$Mpc. 

For both $\xi(s)$ and $P(k)$, we consider intervals of $\Delta \alpha = 0.001$ in the range $0.8 < \alpha < 1.2$ and find the minimum $\chi^2$ value at each $\alpha$ when varying the bias and shape parameters. For the correlation function, we set $\Sigma_{nl} = 8h^{-1}$Mpc for the standard case and $\Sigma_{nl} = 4h^{-1}$Mpc in the reconstructed case (as in \citealt{Xu12mon,alph}). For $P(k)$, $\Sigma_{nl}$ is allowed to vary within a Gaussian prior defined by 8$\pm2 h^{-1}$Mpc. Allowing some freedom in the damping term is important for the $P(k)$ fit, where the BAO signature in the damped high-$k$ tail is of direct importance. We then determine the likelihood distribution of $\alpha$ assuming $p(\alpha) \propto e^{-\chi^2(\alpha)/2}$ and that the total probability integrates to 1 (interpolating to find the likelihood at any given $\alpha$ value).  

\subsection{Physical Implications of Anisotropic Clustering}
\subsubsection{Structure Growth}
\label{sec:SGmod}
At large scales, the amplitude of the velocity field is given by the rate of change of the linear growth rate, $f \equiv {\rm d log} D/{\rm d log} a$, where $D$ is the linear growth rate and $a$ is the scale factor of the Universe. Assuming General Relativity, the value of $f$ is a deterministic function of $\Omega_m(z)$, and thus measurements of the velocity field test the validity of General Relativity on cosmological scales.

When clustering is measured in redshift space, peculiar velocities cause
distortions to the true separation between galaxies. For the power spectrum,
this results in an enhancement of the clustering in the line-of-sight (LOS) direction and in
linear theory can be described as related to the power in the velocity field \citep{Kaiser87}:
\begin{equation}
  P_{\rm g}^{\rm s}(k,\mu_k) = (b+f\mu^2_k)^2P^r_m(k),
\end{equation}
where $b$ is the galaxy bias relating the amplitude of the galaxy field to the
matter field, $\mu_k$ is the {\rm Cosine} of the angle between $\vec{k}$ and the LOS, and $P^r_m(k)$ is the real-space matter power spectrum. In the configuration
space linear theory predicts 
\begin{eqnarray}
\xi_0(b,f) & = &\xi_{M}\left(b^2+2/3bf+1/5f^2\right)\label{eq:m0}\\
  \xi_2(b,f) &=& \left(\frac{4}{3}bf+\frac{4}{7}f^2\right)[\xi_{M}-\xi_{M}'], \label{eq:m2}
\label{eq:m4}
\end{eqnarray}
\noindent
where $\xi_{M}$ is the real-space matter correlation function and
\begin{equation}
  \xi'\equiv3s^{-3}\int^s_0\xi(r')r'^2dr' 
\end{equation}
\citep{Ham92}. The quantity $\xi_{\ell}$ is the multipole moment of the redshift space correlation function given by
\begin{equation}
\xi_{\ell}(s) =  \frac{(2\ell + 1)}{2}\int_{-1}^{1} {\rm d}\mu L_{\ell}(\mu)\xi(s,\mu),
\label{eq:ximult}
\end{equation}
where $L_{\ell}$ denotes the $\ell$th-order Legendre polynomial. Given $\xi_{M}$, one can measure $b$ and $f$ from measurements of $\xi_{0,2}$. However, both measurements depend on the amplitude of $\xi_{M}$, parameterized by $\sigma_8(z)$; \cite{PW09} and \cite{SP09} have shown that measuring the amplitude of $\xi_{0,2}$ in terms of $f\sigma_8(z),b\sigma_8(z)$ allows tests of Dark Energy and General Relativity.

A more accurate model (than the linear theory prediction described above) of the anisotropic correlation function,
$\xi(r_{\sigma},r_{\pi})$, was developed in \cite{RW11}. Here $r_{\sigma}$ is
the transverse separation, and $r_{\pi}$ is the LOS separation measured in
redshift space. This model has been used in \cite{Reid12RSD}, and \cite{SamDR9,Samushia13} to model the anisotropic correlation function,
$\xi(r_{\sigma},r_{\pi})$. \cite{RW11} compared the measurements from $N$-body
simulations with the predictions of their streaming model in which 
\begin{equation}
1+\xi(s_{\sigma},s_{\pi}) = \int \left[1+\xi^r(r)\right]e^{\left(\frac{-[s_{\pi}-y-\mu v_{12}(r)]^2}{2\sigma^2_{12}(r,\mu)}\right)}\frac{dy}{\sqrt{2\pi\sigma^2_{12}(r,\mu)}},
\label{eq:xiRSD}
\end{equation}
where $\xi^r(r)$ is the real-space correlation function, $y$ is the real-space
LOS separation, $\mu$ is $y/r$, $v_{12}$ is the mean infall velocity of galaxies
with real-space separation $r$, and $\sigma^2_{12}(r,\mu)$ is the rms dispersion
of the pairwise velocity. The terms $\xi^r(r)$,
$\sigma^2_{12}(r,\mu)$, and $v_{12}(r)$ can all be computed using perturbation
theory frameworks, given the real-space host halo bias. \cite{Reid12RSD} also
added a nuisance term $\sigma^2_{FoG}$, which adds an isotropic velocity
dispersion that accounts for the motion of galaxies within halos. Eq.
\ref{eq:xiRSD} can be combined with Eq. \ref{eq:ximult} to obtain theoretical
predictions for $\xi_{\ell}(s)$. \cite{RW11} demonstrated that for the halo population
with $b\sim 2$ (the DR10 CMASS full sample has $b\sim2$) the model matches
$N$-body simulation measurements of $\xi_0$ and $\xi_2$ to within 2 per cent. We
apply this model on our Red and Blue samples and their cross-correlation in Section \ref{sec:RSD}.

\subsubsection{Distance Scale Information}
Additional information is gained by considering the anisotropic effect on the
measured clustering induced by the difference between the true geometry and that
assumed when the clustering is measured; this is known as the
Alcock-Paczynski (AP) test \citep{AP79}. Similar to the case of measuring the BAO scale, we
consider the effect as a scaling of the true separation, but we now consider the
scaling perpendicular, $\alpha_{\perp}$, and parallel, $\alpha_{||}$, to the
LOS, where 
\begin{equation} \xi^{fid}(r_{\sigma},r_{\pi}) =
  \xi^{true}(\alpha_{\perp}r_{\sigma},\alpha_{||}r_{\pi}) \end{equation}
and
\begin{equation}
\alpha_{\perp} = \frac{D^{\rm true}_A(z)}{D^{\rm fid}_A(z)}, ~\alpha_{||} =
\frac{H^{\rm fid}(z)}{H^{\rm true}(z)}.
\end{equation}
\noindent
These two quantities are related to the parameter $\alpha$ defined in Section~\ref{sec:baomod} by 
\begin{equation}
  \alpha = \left(\alpha_\perp^2\alpha_{||}\right)^{1/3}.
\end{equation}
As explained in the previous section, measurement of $\xi_0$ constrains the
distance scale $D_V(z)$. Considering the AP effect, $\xi_2$ allows measurement
of 
\begin{equation} 
  F(z) = (1+z)D_A(z)H(z)/c.  
\end{equation}
\noindent
Thus, measurements of $\xi_{0,2}$ allow one to break the degeneracy between
$D_A$ and $H$. In Section \ref{sec:RSD}, we
present measurements of $\alpha_{\perp}, \alpha_{||}$ from our Red and Blue
samples. Following \cite{Anderson13}, we define
\begin{equation}
  \epsilon = \left(\frac{\alpha_\perp}{\alpha_{||}} - 1\right)^{1/3}.
  \label{eq:ep}
\end{equation}
These measurements of $\alpha_{\perp}$ and $\alpha_{||}$ contain
information from the shape of the $\xi_{0,2}$ and therefore will not be
identical to the similar measurements derived from the BAO fitting described in Section \ref{sec:baomod}. 

When comparing Eq. \ref{eq:xiRSD} to the clustering we measure, we assume that
the clustering of our galaxy populations can be modelled as having a single host
halo bias. Strictly speaking, this is an approximation, but we verify this assumption does
not bias our results when applied to our mock catalogs in Section \ref{sec:RSD}.
The effective bias of our Blue sample is $b \sim 1.6$; \cite{RW11} show that the RSD model that we use
is expected to be less accurate in that bias range. This means that our estimates of $f\sigma_8$
are less robust than the ones derived from the full sample, if used for the
purposes of deriving cosmological constraints. However, for the
purpose of testing the possible systematic effects coming from different galaxy
populations, which is the main goal of our paper, the accuracy of the model is
acceptable.

We use the same fitting method as in \cite{Reid12RSD} and \cite{Samushia13}: We marginalize over
parameters describing the shape of the linear power-spectrum and the FOG
velocity dispersion to obtain constraints on $b\sigma_8(z)$, $f\sigma_8(z)$,
$\alpha_{\perp}$, and $\alpha_{||}$. Compared to methods that fit for the BAO scale information only, e.g., \cite{Anderson13},
we use a considerable amount of the information contained in the full shape of the $\xi_{0,2}$ measurements.

\cite{McDSel09} demonstrate that using two samples with different bias that
occupy the same volume may allow improved measurements of $f\sigma_8(z)$, due to
the fact that each sample will trace the same underlying density field and
therefore have highly correlated cosmic variance. This feature allows one to measure the
relative clustering amplitude with considerably lower cosmic variance
uncertainty than if the sample occupied separate volumes, and thus obtain lower
uncertainty on $f\sigma_8(z)$. We will investigate whether one can improve BOSS
$f\sigma_8$ measurements using this technique in Section \ref{sec:2t}.

\section{Clustering Measurements}
\label{sec:meas}
\subsection{Calculating Clustering Statistics}

We calculate the correlation function as a function of the redshift-space separation $s$ and the Cosine of the angle to the line-of-sight, $\mu$,
 using the standard \cite{LS} method 
\begin{equation}
\xi(s,\mu) =\frac{D_1D_2(s,\mu)-D_2R_1(s,\mu)-D_1R_2(s,\mu)+R_1R_2(s,\mu)}{R_1R_2(s,\mu)}, 
\label{eq:xicalc}
\end{equation}
where $D$ represents the galaxy sample and $R$ represents the uniform random sample that simulates the selection function of the galaxies. $DD(s,\mu)$ thus represent the number of pairs of galaxies with separation $s$ and orientation $\mu$. We use at least fifty times the number of galaxies in each of our random samples.

We calculate $\xi(s,|\mu|)$ in evenly-spaced bins of width 4 $h^{-1}$Mpc in $s$ and 0.01 in $|\mu|$. We then determine the first two even moments of the redshift-space correlation function via 
\begin{equation}
\frac{2\xi_{\ell}(s)}{2\ell+1} = \sum^{100}_{i=1} 0.01\xi(s,\mu_i)L_{\ell}(\mu_i), 
\label{eq:xiell}
\end{equation}
where $\mu_i = 0.01i-0.005$.

As described in Section \ref{sec:wsys}, we apply a weight to each galaxy, $w_{\rm sys}$, that corrects for the systematic relationships we are able to identify. In addition, we weight both galaxies and randoms based on the number density as a function of redshift \citep{FKP}, via
\begin{equation}
w_{\rm FKP}(z) = \frac{1}{1+P_{\rm FKP}n(z)},
\label{eq:wfkp}
\end{equation}
where we set $P_{\rm FKP} = 20000 h^{-3}$Mpc$^{3}$, the same as is used for the full CMASS sample in \cite{alph}. For our $\xi(s)$ BAO measurements, we find that applying the $w_{\rm FKP}$ weights reduces the mean recovered uncertainty by 3 per cent for the Red sample and 2 per cent for the Blue sample. Given the size of these improvements, optimizing the weights by accounting for the difference in clustering amplitude of the two samples is unlikely to significantly increase the precision of our results.

The total weight applied to each galaxy is
\begin{equation}
w_{\rm tot} = w_{\rm FKP}w_{\rm sys}
\end{equation}
and for each random point $w_{\rm tot} = w_{\rm FKP}$. Following \cite{Ross12}, we assign redshifts to our random samples by randomly selecting them from the galaxy samples. These procedures, except for the inclusion of weights for airmass, match the treatment applied to the full CMASS sample in \cite{Aardwolf}.  

We measure the spherically-averaged power spectrum, $P(k)$, using the standard Fourier technique of \cite{FKP}, as described in \cite{ReidDR7} and \cite{alph}. We calculate the spherically-averaged power in $k$ bands of width $\Delta k = 0.008h$Mpc$^{-1}$ using a 2048$^3$ grid. The weights are taken into account by using the sum of $w_{tot}$ over the galaxies/randoms at each gridpoint. 

\subsection{Reconstruction}
\label{sec:rec}

The information in the observed galaxy density field can be used to estimate the matter density field and thus the displacement vector (away from the primordial position) of each galaxy. Moving the galaxies backwards along their estimated displacement vectors results in a ``reconstructed'' galaxy field with a BAO feature that is less degraded by non-linear structure formation \citep{Eis07rec,PWC09}. It has been demonstrated that such a technique can significantly improve BAO measurements when applied to observed galaxy samples \citep{Pad12,Aardwolf}. 

The reconstruction algorithm we apply is developed adopting the methods outlined in \cite{Eis07rec} and \cite{Pad12}. We use the full CMASS galaxy sample (and mock galaxy samples) to produce a smoothed galaxy over-density field. We apply a Gaussian smoothing with scale $15h^{-1}$Mpc (as used in \citealt{alph,Aardwolf}), appropriate to ensure that only regions important for the BAO signal degradation are included \citep{Eis07rec}. Using this smoothed galaxy over-density field and assuming the CMASS galaxy field is biased with respect to the matter field with $b=1.85$, the Lagrangian displacement field $\bold{\Psi}$ is approximated to first order using the Zel'dovich approximation \citep{Zel70}. We are using the full CMASS sample to estimate the displacement field.

\begin{figure}
\includegraphics[width=84mm]{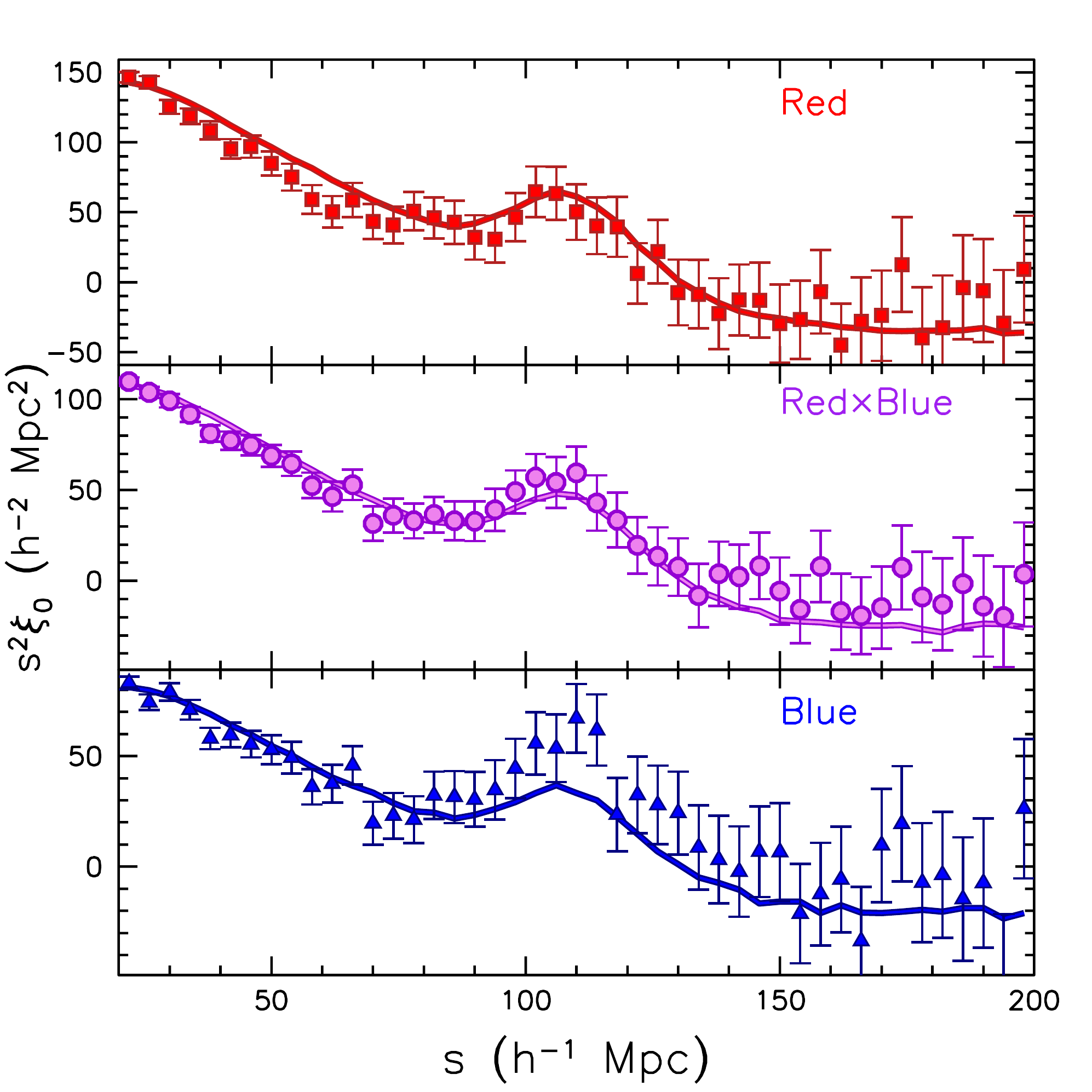}
  \caption{The measured spherically-averaged correlation function, $\xi_0$, for the Red (red points; top) and Blue (blue points; bottom) data samples, and their cross-correlation (purple points; middle). The smooth curves in each panel display the mean $\xi_0$ of the 600 mock realizations of each respective sample. Each mean is a good match to the $\xi_0$ of the data, as $\chi^2/{\rm dof} < 1$ for each.}
  \label{fig:redbluemock}
\end{figure}
\begin{figure}
\includegraphics[width=84mm]{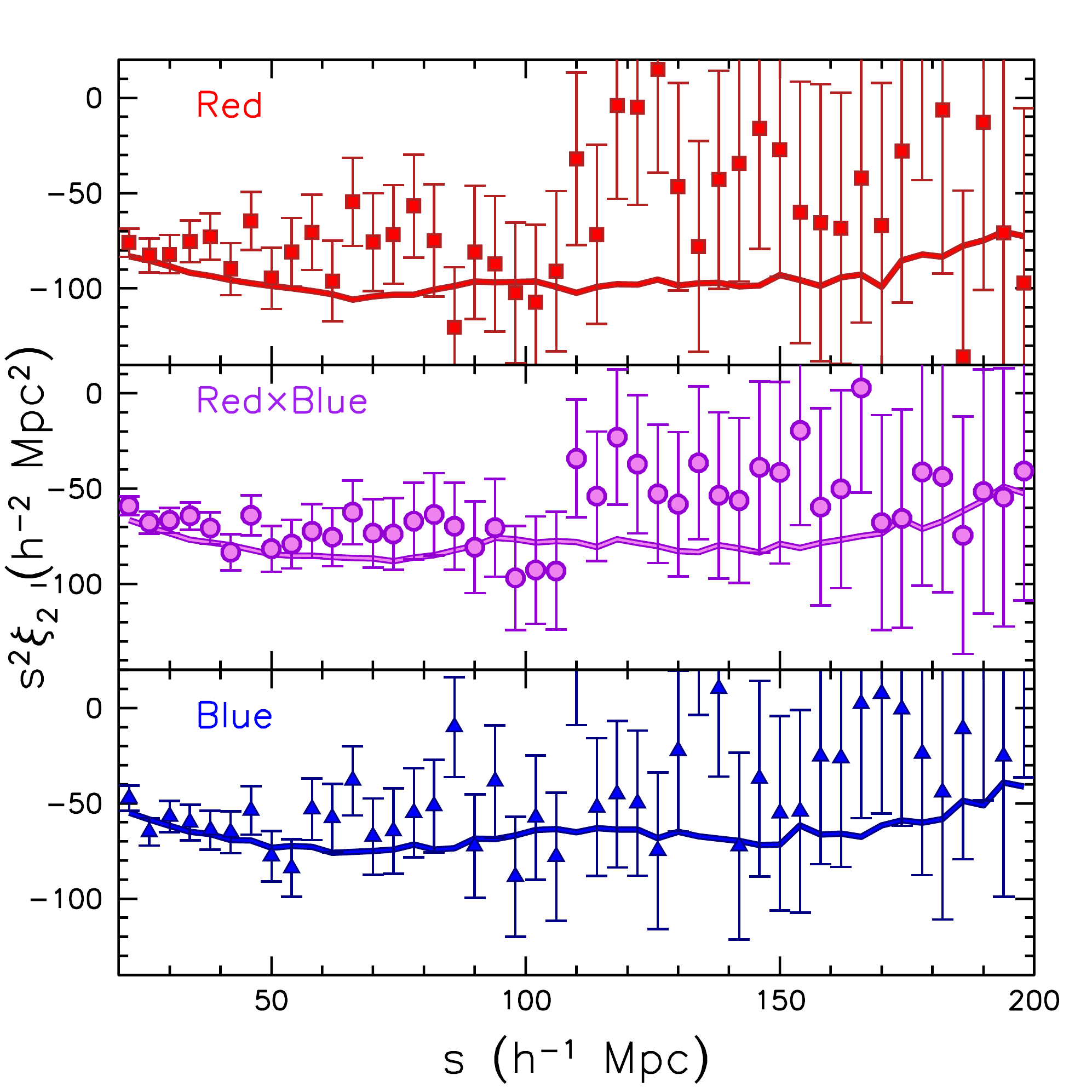}
  \caption{Same as Fig. \ref{fig:redbluemock}, but for the measured quadropole of the correlation function, $\xi_2$. Each mean is a good match to the $\xi_2$ of the data, as $\chi^2/{\rm dof} < 1$ for each.}
  \label{fig:redbluemock2}
\end{figure}

For both the Red and Blue samples, we move the galaxies and random points back along the displacement vectors. For the galaxies, we included an additional $-f\left(\bold{{\Psi}\cdot \hat{s}}\right)\bold{\hat{s}}$ shift to remove redshift space distortions, where $f = 0.744$ is the amplitude of the velocity field (defined in Section \ref{sec:SGmod}) in our assumed cosmology and $\bold{\hat{s}}$ points along the radial direction. The reconstructed field is constructed from the displaced galaxy field minus the shifted random field. For the post-reconstruction field, the correlation function is given by \citep{Pad12}
\begin{equation}
\xi^{\rm rec}(s,\mu) =\frac{D_sD_s(s,\mu)-2D_sS(s,\mu)+SS(s,\mu)}{RR(s,\mu)},
\label{eq:xicalc}
\end{equation}
where $D_s$ is the shifted galaxy field, $S$ is the shifted random field and $R$ is the original random field.

\subsection{Comparison Between Real and Mock Clustering}
\label{sec:RMcom}
Fig. \ref{fig:redbluemock} displays the measured $\xi_0$ for each sample; the Red sample displayed in red in the top panel,
the Blue sample displayed in blue in the bottom panel and their cross-correlation displayed in purple in the bottom panel. The smooth curves display
the mean of the $\xi_0$ calculated from 600 mock realizations of each sample and the error-bars are their standard deviation. The $\xi_0$ measured from the CMASS data agree well with the mean mock
$\xi_0$; we compare the 45 data points in the range $20 < s < 200$ to the mean of the mock $\xi_0$ and we find $\chi^2 = 37.5, 44.7,
30.9$ for the Red, Blue, and cross-correlation monopoles, each therefore having a $\chi^2/{\rm dof} < 1$.

Fitting our fiducial $\xi_{\rm BAO}$ template in the range $20 < s < 200$ and accounting for the growth factor and the value of $f$ (both calculated using our fiducial assumed cosmology) at the $z^{eff}$ (defined by Eq. \ref{eq:zeff}) of each sample, we find a real-space bias of 2.30$\pm$0.09 for the Red sample ($\chi^2_{min} = 33.2$ for 44 dof) and 1.65$\pm$0.07 for the Blue sample ($\chi^2_{min} = 46.5$). Fitting the cross-correlation yields a bias of 1.96$\pm$0.05 ($\chi^2_{min} = 28.0$), consistent with the geometric mean of the measured bias of the two samples (1.95).

Fig. \ref{fig:redbluemock2} displays the same information as Fig. \ref{fig:redbluemock}, but for $\xi_2$. All of the $\xi_2$ measurements are reasonably well fit by the average of the mocks between $21 < s < 199$ (45 data points): the $\chi^2$ are 39.2, 44.4, and 28.4 for the Red, Blue, and Red$\times$Blue $\xi_2$ measurements, once more, each has $\chi^2/{\rm dof} < 1$.

\begin{figure}
\includegraphics[width=84mm]{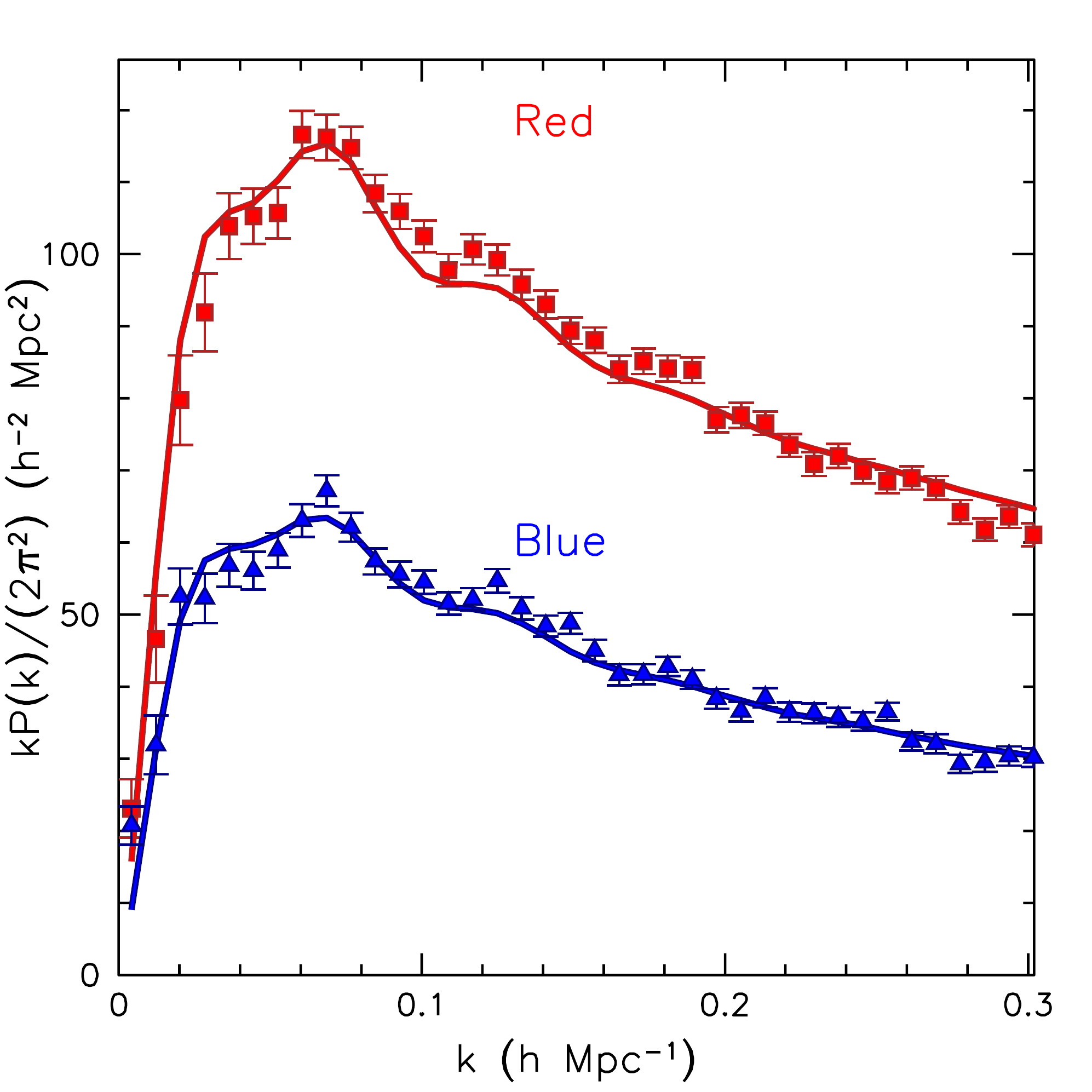}
  \caption{The measured $P(k)$ of the Red (red) and Blue (blue) data samples (points with error-bars) compared to the mean of the $P(k)$ determined from 600 realizations of the respective mock samples (smooth curves). The $\chi^2$ are slightly high, as for the 35 data points with $0.02 < k < 0.3h$Mpc$^{-1}$, $\chi^2_{Red}=65$ and $\chi^2_{Blue}=49$.}
  \label{fig:pkplotrb}
\end{figure}

\begin{figure}
\resizebox{84mm}{!}{\includegraphics{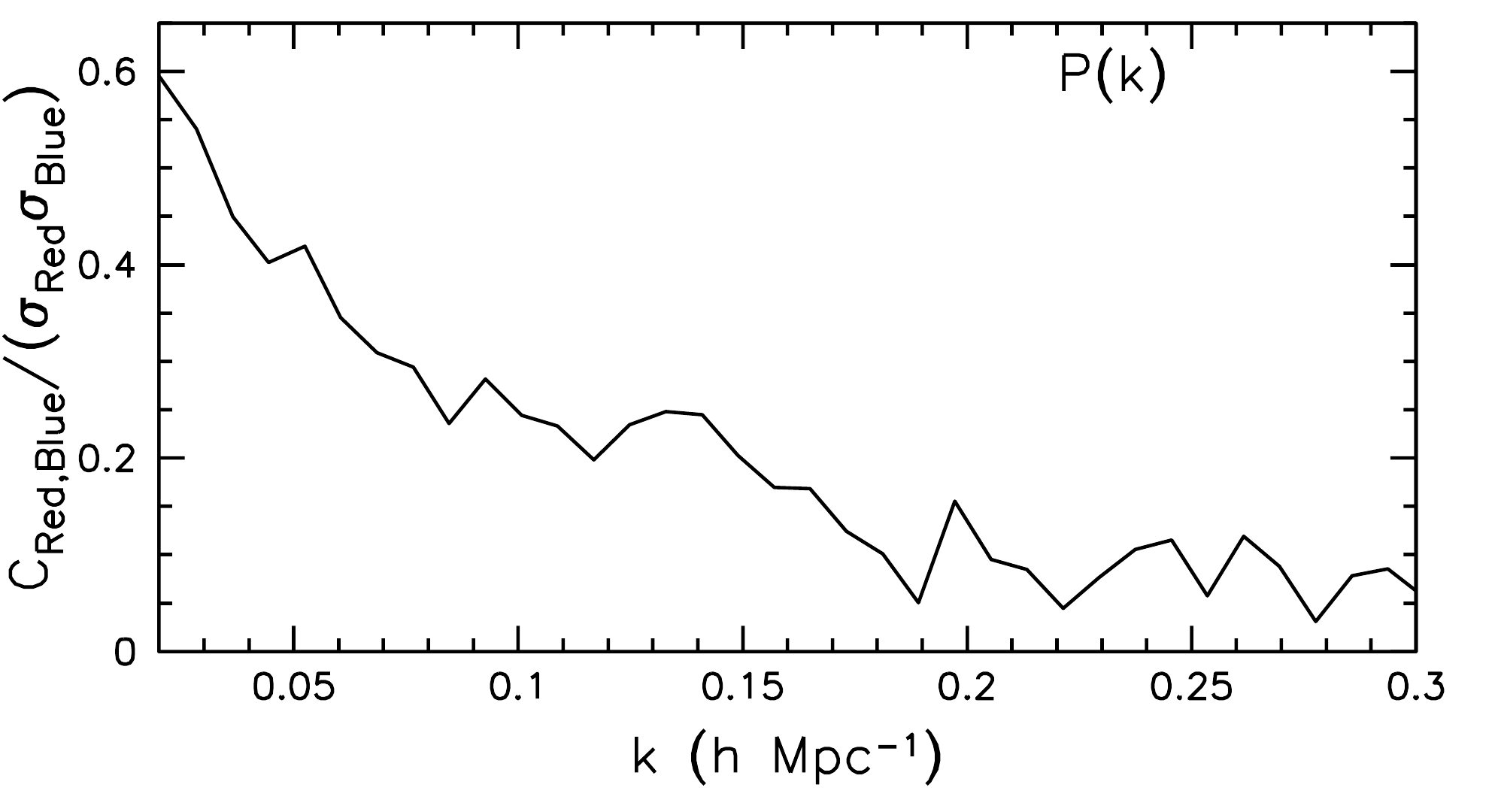}}
\resizebox{84mm}{!}{\includegraphics{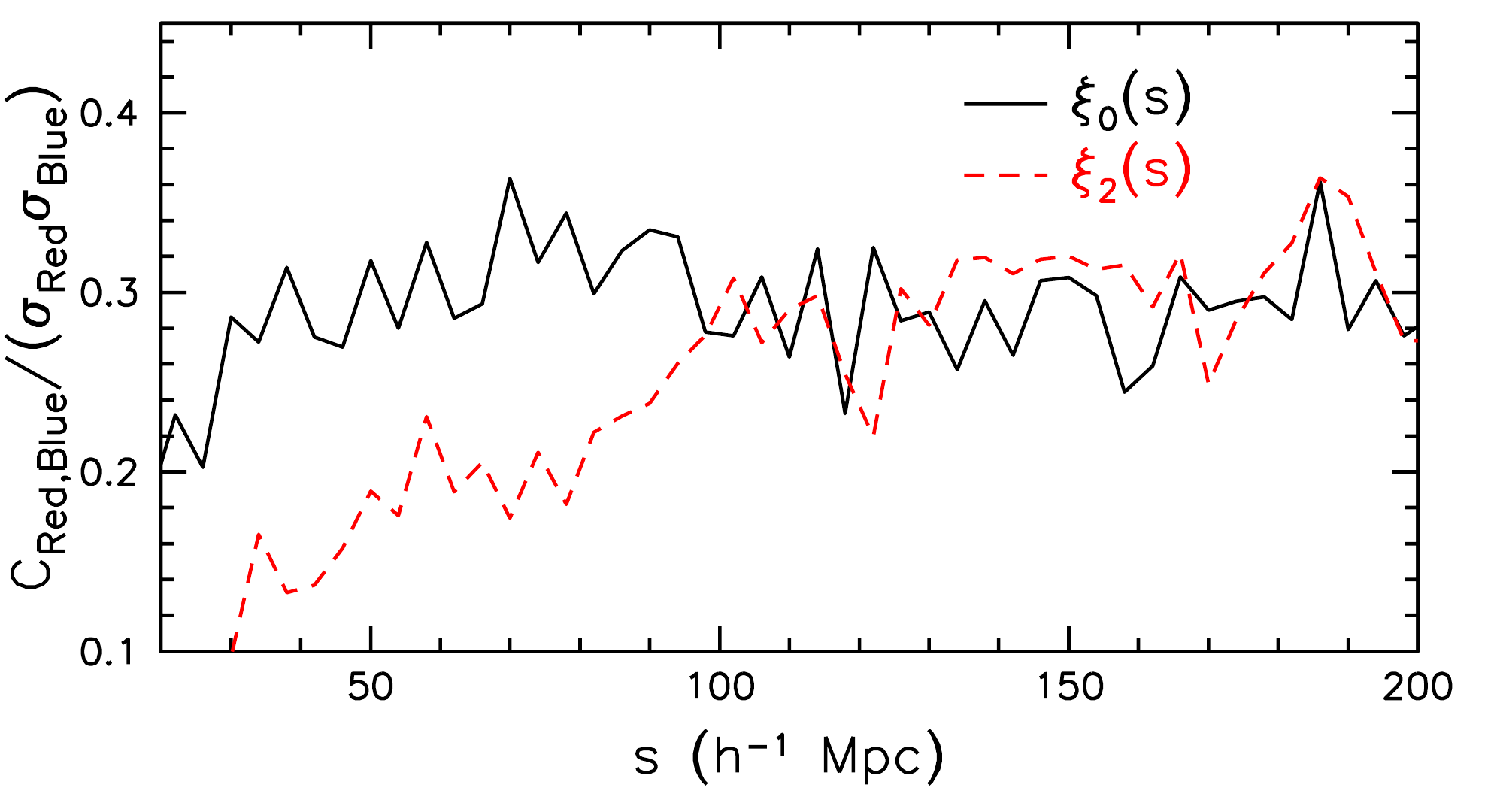}}
  \caption{The correlation between the clustering of 600 mock realizations of the Red and Blue galaxy samples, measured in Fourier space (top $P(k)$) and redshift space (bottom, $\xi(s)$). The correlation changes less as a function of scale for $\xi(s)$ due to the fact that there is significant covariance across measurements in bins of $s$.}
  \label{fig:covredblue}
\end{figure}

Fig. \ref{fig:pkplotrb} displays the measured spherically averaged $P(k)$ for each sample, with red representing the Red sample and blue representing the Blue sample; the points display the measurements determined using CMASS data, the smooth curves represent the mean measurement from 600 mock realizations of each sample, and the error-bars are the standard deviation of the mock $P(k)$ measurements. The amplitudes of the mean mock $P(k)$ appear to be a good match to the CMASS measurements. However, the shape is not a perfect match, as in the range $0.02 < k < 0.3 ~h$Mpc$^{-1}$ (the same 35 data points as are used for the BAO fits) we obtain $\chi^2 = 65$ for the Red sample and $\chi^2=49$ for the Blue. The minimum $\chi^2$ values improve by $\Delta \chi^2 < 2$ when we allow the amplitude of the mean of the mock $P(k)$ to be re-scaled by a constant factor. Thus, it is the mismatch between the respective shapes that causes the poor $\chi^2$. In the following section, we will find reasonable $\chi^2$ values when fitting the BAO position and allowing the smooth shape to be free. The agreement is better for $0.02 < k < 0.1 ~h$Mpc$^{-1}$ (10 data points), in which case we find $\chi^2 = 14$ for the Red sample and $\chi^2 = 11$ for the Blue. A mismatch in the shapes of the mock and CMASS $P(k)$ could be caused by, e.g., the true cosmology differing from the assumed one.

In Fig. \ref{fig:covredblue} we display the correlation between the Red and Blue clustering measurements, as determined from the 600 mock realizations of the respective samples. The $P(k)$ results, displayed in the top panel, show a strong scale dependence. At large scales (small $k$), cosmic variance the dominates the uncertainty. The Red and Blue samples occupy the same volume and are thus strongly correlated at large scales. At small scales (large $k$), the dominant component of the uncertainty is the shot-noise and thus the Red and Blue $P(k)$ are less correlated. For $\xi(s)$ (bottom panel) the measurements in each $s$ bin are strongly covariant. Thus, there is less scale dependence in the correlation between the Red and Blue $\xi(s)$.

\begin{figure}
\includegraphics[width=84mm]{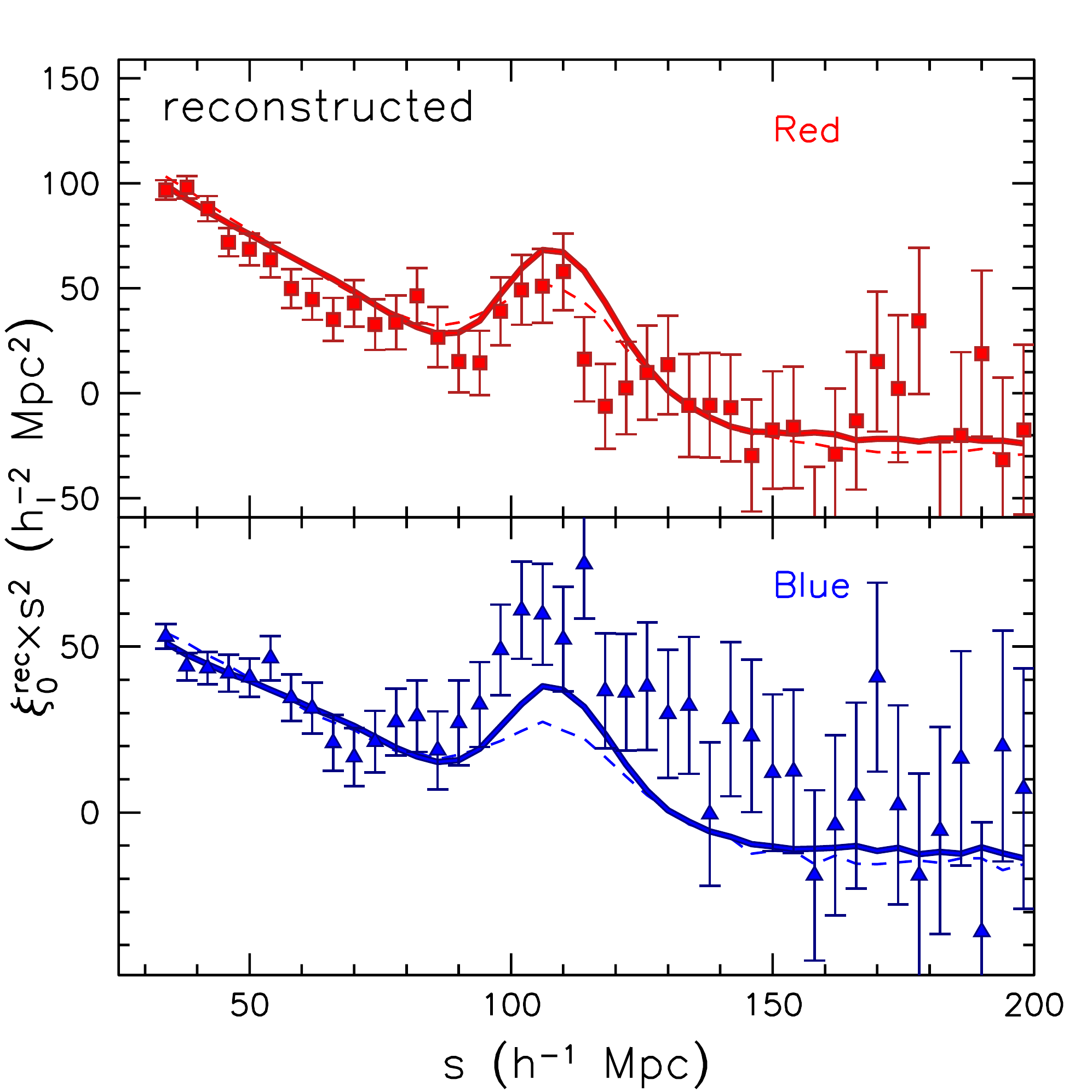}
  \caption{The measured spherically-averaged correlation function, $\xi^{\rm Rec}_0$, after reconstruction, for the Red (red points; top) and Blue (blue points; bottom) samples. The solid curves in each panel are the mean $\xi^{\rm Rec}_0$ of the 600 mock realizations of each respective sample. The lighter dashed curves display the mean $\xi_0$ determined from the un-reconstructed mock realizations of each respective sample, multiplied by a factor that removes the boost in amplitude from RSD (see Eq. \ref{eq:m0}). By eye, the Blue data appears to be a worse fit to the mean of the mocks, but in fact we find $\chi^2_{Blue}=38$ and $\chi^2_{Red}=61$ for the 42 data points in the range $32 < s < 200h^{-1}$Mpc.}
  \label{fig:redbluemockrec}
\end{figure}

Fig. \ref{fig:redbluemockrec} displays the measured $\xi^{\rm Rec}_0$ for the reconstructed CMASS data samples (points with error-bars) compared to the mean $\xi^{\rm Rec}_0$ calculated from 600 reconstructed mock realizations of each sample, with the Red sample results shown in the top panel and the Blue sample results shown in the bottom panel. For the Blue sample, $\chi^2=38.0$ when comparing the CMASS measurements to the mean of the mocks for the 42 data points in the range $32 < s < 200h^{-1}$Mpc. In this same range, $\chi^2=60.6$ for the Red sample. The poor fit for the Red sample is due mainly to the measurements at the largest scales, as for $32 < s < 150h^{-1}$Mpc (29 data points) $\chi^2=35.9$. 

In Fig. \ref{fig:redbluemockrec}, we also display the mean of the un-reconstructed mock sample $\xi_0$ measurements with dashed curves. We have multiplied these curves by a factor $\frac{b^2}{b^2+2/3bf+1/5f^2}$. This factor accounts for the boost in $\xi_0$ amplitude imparted by RSD, as given by Eq. \ref{eq:m0}. The reconstruction algorithm removes this large-scale RSD effect and therefore the amplitude of the pre- and post-reconstruction $\xi_0$ agree after applying this factor. This agreement occurs even though the bias of the full sample ($b=1.85$) is input into the reconstruction algorithm in order to obtain the displacement field. This result implies that, as expected in linear theory, the reconstruction algorithm is correctly identifying the local bias of the Red and Blue fields imbedded in the overall CMASS field.

\section{BAO Measurements}
\label{sec:bao}

\begin{table*}
\begin{minipage}{7in}
\caption{The statistics of BAO scale measurements recovered from the mock and data Red and Blue galaxy samples. The parameter $\langle \alpha \rangle$ is the mean $\alpha$ value determined from 600 mock realizations of each sample, $S_{\alpha} = \sqrt{\langle(\alpha-\langle\alpha\rangle)^2\rangle}$ is the standard deviation of the best-fit $\alpha$ values, $\langle \sigma \rangle$ is the mean 1 $\sigma$ uncertainty on $\alpha$ recovered from the likelihood distribution of each realization, $\alpha_{\rm KS}$ is the $\alpha$ value that minimizes the $D_n$ value obtained when applying the Kolmogorov-Smirnov test to the distribution of recovered $\alpha$ and $\sigma$ values, $\langle \chi^2 \rangle$ is the mean minimum $\chi^2$ value, and ``CMASS $\alpha$'' is the measurement for the data sample.    }
\begin{tabular}{lcccccccc}
\hline
\hline
Case  &  $\langle \alpha \rangle$ & $S_{\alpha}$ &  $\langle \sigma \rangle$ &  $\alpha_{\rm KS}$  & $D_n, P_{\rm KS}$ & $\langle \chi^2 \rangle$/dof  & non-detections & CMASS $\alpha$, $\chi^2$/dof \\
\hline
$P_{\rm Red}(k)$ & 1.0047 & 0.0287  & 0.0268 & 1.0042 & 0.022, 0.93 & 30/29 & 14 & 0.992$\pm$0.025, 33/29 \\
$\xi_{\rm Red}(s)$ & 1.0019 & 0.0281 & 0.0266 & 1.0023 & 0.016, 0.999 & 37/37 & 19 & 1.010$\pm$0.027, 28/37\\
Rec. $\xi_{\rm Red}(s)$ & 0.9993 & 0.0198 & 0.0202 & 0.9985 & 0.027, 0.78 & 37/37 & 1 & 1.013$\pm$0.020, 51/37\\
$P_{\rm Blue}(k)$ & 1.0016 & 0.0402 & 0.0380 & 1.0013 & 0.020, 0.98 & 30/29 & 15 & 0.999$\pm$0.030, 34/29\\
$\xi_{\rm Blue}(s)$ & 0.9980 & 0.0386 & 0.0372 & 0.9990 & 0.029, 0.72 & 37/37 & 20 & 1.005$\pm$0.031, 37/37\\
Rec. $\xi_{\rm Blue}(s)$ & 0.9994 & 0.0296 & 0.0300 & 0.9992 & 0.031, 0.63 & 37/37 & 6 & 1.008$\pm$0.026, 35/37\\
$\xi_{\times}(s)$ & 1.0017 & 0.0310 & 0.0260 &1.0028 & 0.030, 0.67 & 40/37 & 13 & 1.024$\pm$0.024, 20/37\\
\hline
\label{tab:mockbao}
\end{tabular}
\end{minipage}
\end{table*}

\begin{figure}
\includegraphics[width=84mm]{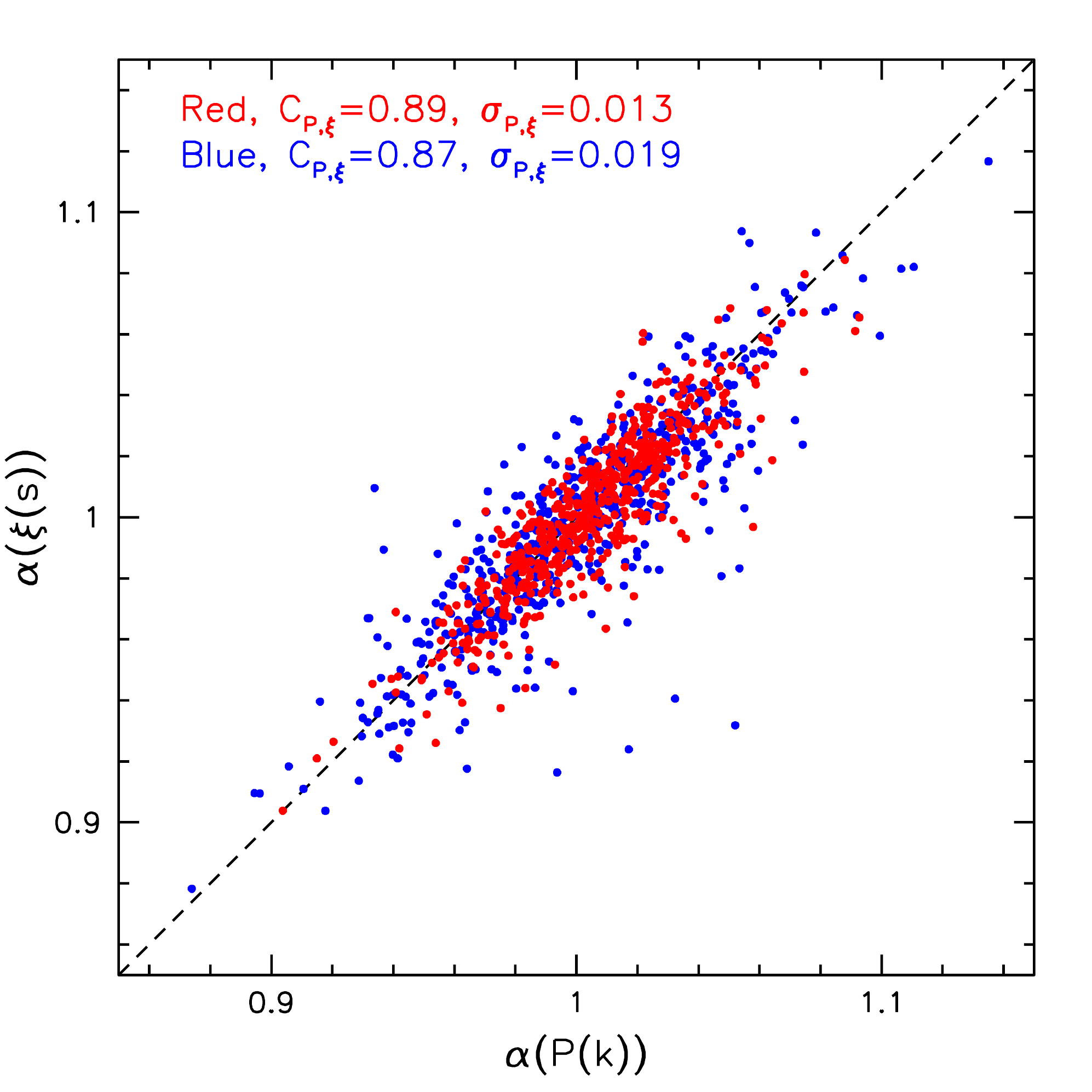}
  \caption{The 600 BAO scale measurements, $\alpha_{\xi}$, recovered from correlation functions of each mock realization vs. the BAO scale, $\alpha_P$, recovered from the power spectrum of the same mock realization, for the Red (red points) and Blue (blue points) samples. The correlation between the two estimates for both the Red and Blue samples is higher than 0.87, as quantified using the $C$ factor defined in Eq. \ref{eq:corr12}, and the mean differences (the labeled $\sigma_{P,\xi}$ values) are both less than 0.35 of the values expected for independent samples.}
  \label{fig:xipkBAO}
\end{figure}

In order to measure the BAO position, as parameterized by the likelihood distribution of $\alpha$, we apply the methodology outlined in Section \ref{sec:baomod} to each of the Blue, Red, and Red$\times$Blue (which we will denote with a subscript $\times$) $\xi_0(s)$ and $P(k)$ measurements determined for the CMASS data samples and the 600 mock realizations of each galaxy sample. We fit $\xi_0(s)$ in the range $30 < s < 200 h^{-1}$Mpc (42 data points) and $P(k)$ in the range $0.02 < k < 0.3 h$Mpc$^{-1}$ (35 data points). 

For each data set we also find the best-fit solution when $P_{\rm BAO}$ is set to zero. Cases where the $\chi^2$ is best when $P_{\rm BAO}$ is set to zero are defined as non-detections. Non-detections happen at worst 3.3\% of time (the $\xi_{0,\rm Blue}(s)$ measurements prior to reconstruction) and at best 0.2\% of the time (for the post-reconstruction Red $\xi_0(s)$ measurements). We exclude non-detections when we determine the mean and variance of the maximum likelihood of $\alpha$ values recovered from the mock samples. The statistical properties of these measurements are summarized in Table \ref{tab:mockbao}. 

For each set of clustering measurements, we have compared the distribution of $(\alpha-\alpha_{\rm KS})/\sigma_{\alpha}$ to a unit normal distribution using the Kolmogorov-Smirnov (KS) test. In order to find the Gaussian distribution most consistent with the distribution of mock results, we found the value of $\alpha_{\rm KS}$ that minimizes the $D_n$ value for each sample. Table \ref{tab:mockbao} summarizes the results of the KS tests. The low $D_n$ and high $P_{\rm KS}$ values suggest our use of the $\chi^2$ statistic to determine the maximum likelihood and 1$\sigma$ uncertainty values is appropriate. As expected, the $\alpha_{\rm KS}$ values are close to the mean $\alpha$ values, but the agreement between the results recovered from $\xi_0(s)$ and  $P(k)$ measurements is better for the $\alpha_{KS}$ values than for the mean $\alpha$ values. The difference in $\alpha_{\rm KS}$ values recovered from the $P(k)$ and $\xi_0(s)$ measurements is 0.002 for both the Red and Blue samples, suggesting a systematic uncertainty of this order.

Prior to reconstruction, a small shift, due to non-linear structure growth, is expected in the BAO position (see, e.g., \citealt{Eis07,Angulo08,Pad09,McCullagh13}). In terms of $\alpha$, \cite{Pad09} predict a shifts of order $0.005D^2(z)$ for samples with $b=2$ and $0.002D^2(z)$ for samples with $b=1$. We find similar behaviour in our mock samples, as the $\alpha_{\rm KS}$ values for the Blue sample are 0.003 smaller than those of the Red sample for both $P(k)$ and $\xi(s)$. The significance of the difference is 2$\sigma$ given the uncertainty on the mean of the 600 realizations (as the uncertainty on the mean is the standard deviation divided by $\sqrt{600}$, $\sim$0.001 for both). Applying reconstruction moves the mean $\alpha$ values closer to 1 and brings the Red and Blue samples into better agreement; both of these results are as expected (\citealt{Pad12,alph}). The significance of the difference between the Red and Blue $\alpha_{\rm KS}$ after applying reconstruction is reduced to less than 1$\sigma$. Expected or not, all of the deviations from 1 we find in the mean $\alpha$ measurements or $\alpha_{\rm KS}$ are negligible ($<0.2\sigma_{\alpha}$) compared to the mean recovered uncertainty, and we cannot expect any to be detectable in our CMASS data samples.

\begin{figure}
\includegraphics[width=84mm]{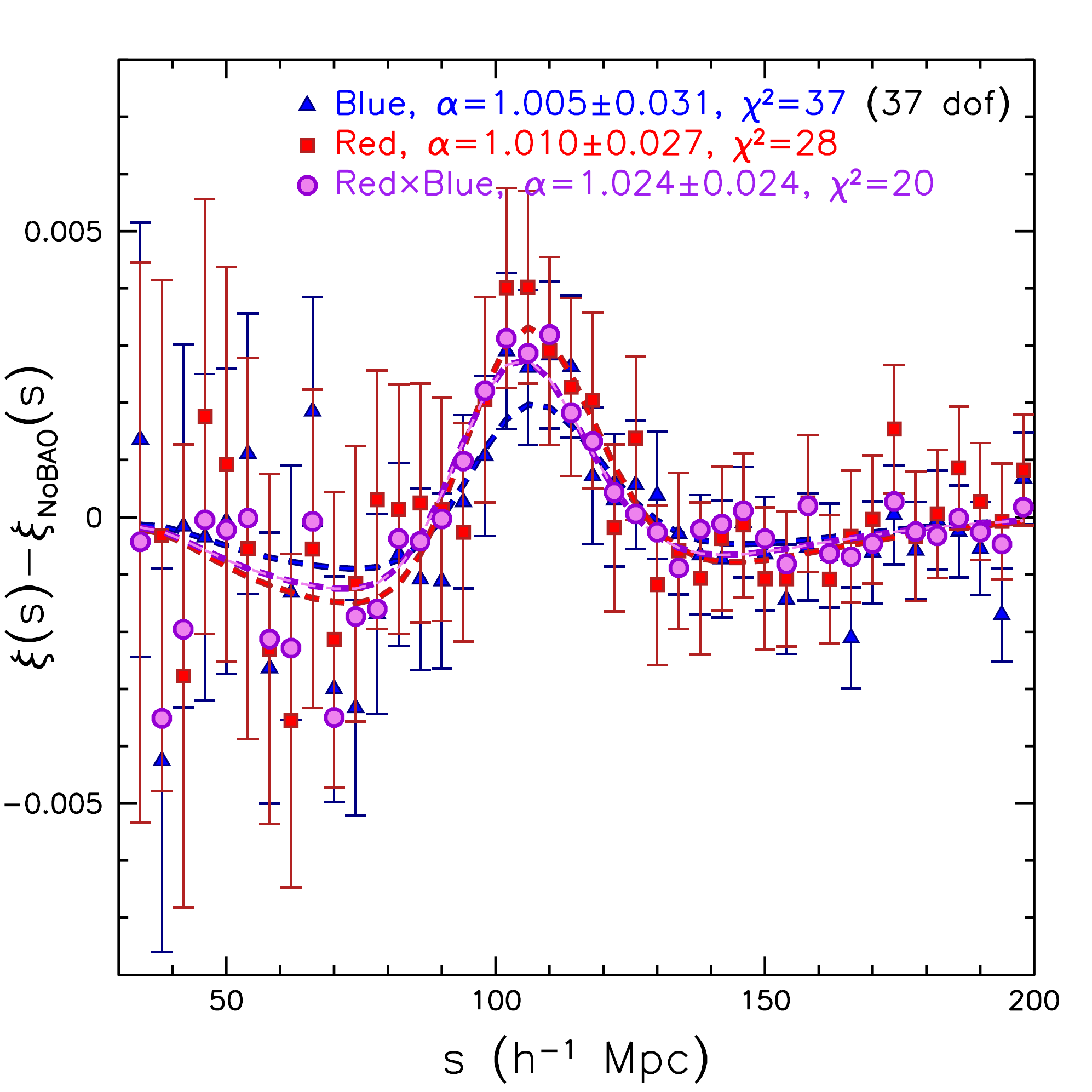}
  \caption{The measured $\xi_0$ (points with error-bars) and the best-fit BAO model (dashed curves) for the Red (red) and Blue (blue) data samples and their cross-correlation (purple). Each has had the smooth component of the best-fit model subtracted. Clear agreement is observed in the location of the BAO peak, and confirmed by the best-fit $\alpha$ values that are labeled. For clarity, we have omitted the error-bars for the cross-correlation.}
  \label{fig:xi0BAOfits}
\end{figure}

\begin{figure}
\includegraphics[width=84mm]{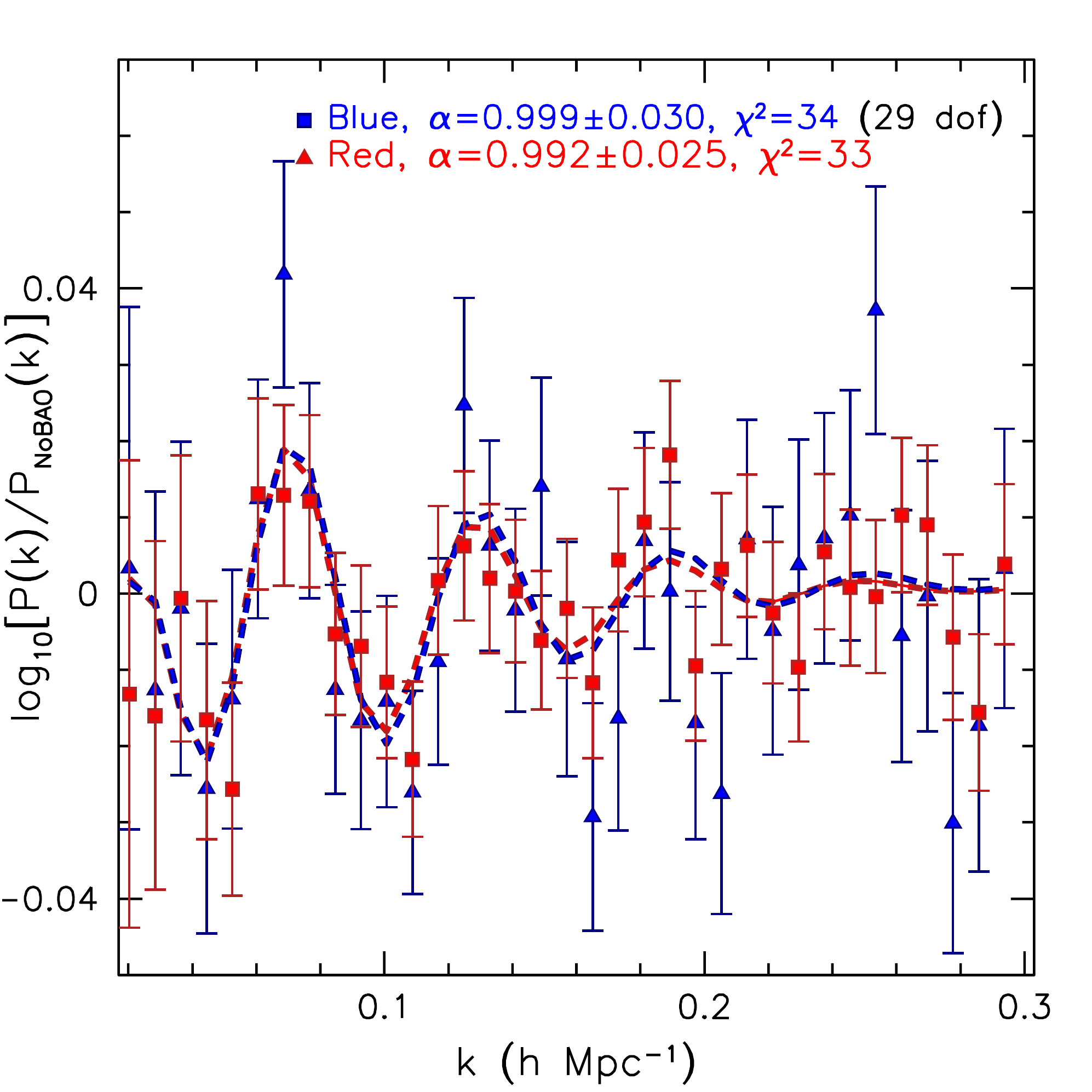}
  \caption{The measured $P(k)$ (points with error-bars) and the best-fit BAO model (dashed curves), both divided by the smooth shape component of the best-fit model, for the Red (red) and Blue (blue) data samples. Clear agreement is observed in the location of the BAO feature, and confirmed by the best-fit $\alpha$ values that are labeled.}
  \label{fig:pkBAOfits}
\end{figure}

The modelling we employ to fit the BAO scale was designed, in part, to maximize the consistency between the measurements obtained from $\xi(s)$ and $P(k)$. Our tests on the mocks confirm that we have achieved a tight correlation. We show the $\alpha$ recovered from $\xi(s)$ versus that recovered for $P(k)$, for both the Red and Blue samples in Fig \ref{fig:xipkBAO}. The correlation, $C_{P,\xi}$, is given by
\begin{equation}
C_{1,2} = \frac{{\rm Cov}_{1,2}}{\sigma_1\sigma_2},
\label{eq:corr12}
\end{equation}
where we use the standard deviation of mock values as $\sigma$. We find $C_{P,\xi} = 0.89$ for the Red sample and $C_{P,\xi} = 0.87$ for the Blue sample. Defining $\sigma_{P,\xi} = \sqrt{\langle (\alpha_P-\alpha_{\xi})^2 \rangle}$, we find $\sigma_{P,\xi}=0.013$ for the Red sample and $\sigma_{P,\xi} = 0.019$ for the Blue. For both datasets, this value is less than 0.35 the dispersion expected for independent samples. 

The correlation between the Red and Blue BAO measurements recovered from the mock realizations is 0.15 for $\xi(s)$ and 0.14 for $P(k)$. These  values are close to the correlation between the Red and Blue $P(k)$ measurements at $k=0.15$, as shown in Fig. \ref{fig:covredblue}. This scale is close to the mid-point of the scales used to fit the $P(k)$ BAO (see Fig. \ref{fig:pkBAOfits}). In Section \ref{sec:RSD} we find a larger correlation (0.37) between the Red and Blue growth measurements, suggesting the effective $k$ for the growth measurements is smaller than for BAO measurements.

Fig. \ref{fig:xi0BAOfits} displays the measured $\xi_0(s)$, using CMASS data, and the best-fit model, both with $\xi_{\rm NoBAO}(s)$ subtracted, for each of the Blue, Red, and Red$\times$Blue measurements. As implied by the agreement displayed in Fig. \ref{fig:xi0BAOfits}, the $\chi^2$ values for the best-fit models are good, as all are smaller than 1 per degree of freedom. The best-fit $\alpha$ values differ by at most 0.014 (between Red$\times$Blue and Red measurements). Quantifying the difference as 
\begin{equation}
d_{\alpha}(1,2) = \left(\frac{(\alpha_{1}-\alpha_{2})^2}{\sigma^2_{1}+\sigma^2_{2}}\right)^{1/2}
\end{equation}
we find that 318 of the mock samples (more than 50 per cent) have a larger $d_{\alpha}(\xi_{Red\times Blue},\xi_{Red})$ than we find for CMASS. The $\alpha$ measurements are clearly consistent. Narrowing the fit range to $50 < s < 160 h^{-1}$Mpc (27 data points) has a negligible effect, as each of the measured $\alpha$ values change by less than 0.1$\sigma$.

The uncertainties we recover from the CMASS data $\xi_0(s)$ BAO measurements are typical of those recovered from the mock realizations. The uncertainty on the Blue data sample measurement (0.031) is better than the mean uncertainty recovered from the Blue mock realizations (0.037). However, we find that 147 of the mock Blue $\xi_0(s)$ (24.5 per cent) yield lower uncertainty, suggesting the Blue data value is not unusual. The uncertainty we recover from the Red data $\xi_0(s)$ BAO measurement (0.027) matches the mean uncertainty we find for the mock samples. The uncertainty we find for the BAO scale measured from the cross correlation of the Red and Blue data samples is typical, as we find 205 of the mock realizations (34 per cent) yield an uncertainty lower than 0.024.

Fig \ref{fig:pkBAOfits} displays the measured $P(k)$ and the best-fit BAO model for the Blue and Red data samples, all divided by the $P_{\rm NoBAO}$ component of the best-fit model. The best-fit measurements appear to agree well, and this is confirmed by $\chi^2$ values that are less than 1.2/dof. The Red and Blue BAO measurements are clearly consistent with each other, as they differ by only 0.007. Narrowing the fit range to $0.04 < k < 0.2 h$Mpc$^{-1}$ (20 data points) has a negligible effect, as each of the $\alpha$ values change by less than 0.1$\sigma$. Similar to the $\xi_0(s)$ result, the uncertainty on the CMASS data Blue BAO measurement (0.030) is better than the mean uncertainty recovered from the mock realizations (0.038), but we find 126 mock Blue $P(k)$ measurements (21 per cent) that yield $\sigma_{\alpha} < 0.030$. 

The power spectrum and correlation function BAO measurements are clearly consistent for the Blue data sample. We find $\alpha_{P,Blue} = 0.999\pm0.030$, $\alpha_{\xi,Blue} = 1.005\pm0.031$ and the mean difference, $\langle |\alpha_{\xi,Blue}-\alpha_{P,Blue}|\rangle$, recovered from Blue mock realizations is 0.019 and the majority of these realizations have a larger $d_{\alpha}$ value. We find a larger discrepancy for the Red data sample ($\alpha_{P,Red} = 0.992\pm0.025$, $\alpha_{\xi,Red} = 1.010\pm0.027$), and the difference is larger than the mean difference we find in the mock samples, 0.013. However, for 61 of the Red mock realizations we find a larger $d_{\alpha}(P_{Red},\xi_{Red})$ than we find for our data sample, and thus the chance of finding such a difference was greater than 10 per cent.

\begin{figure}
\includegraphics[width=84mm]{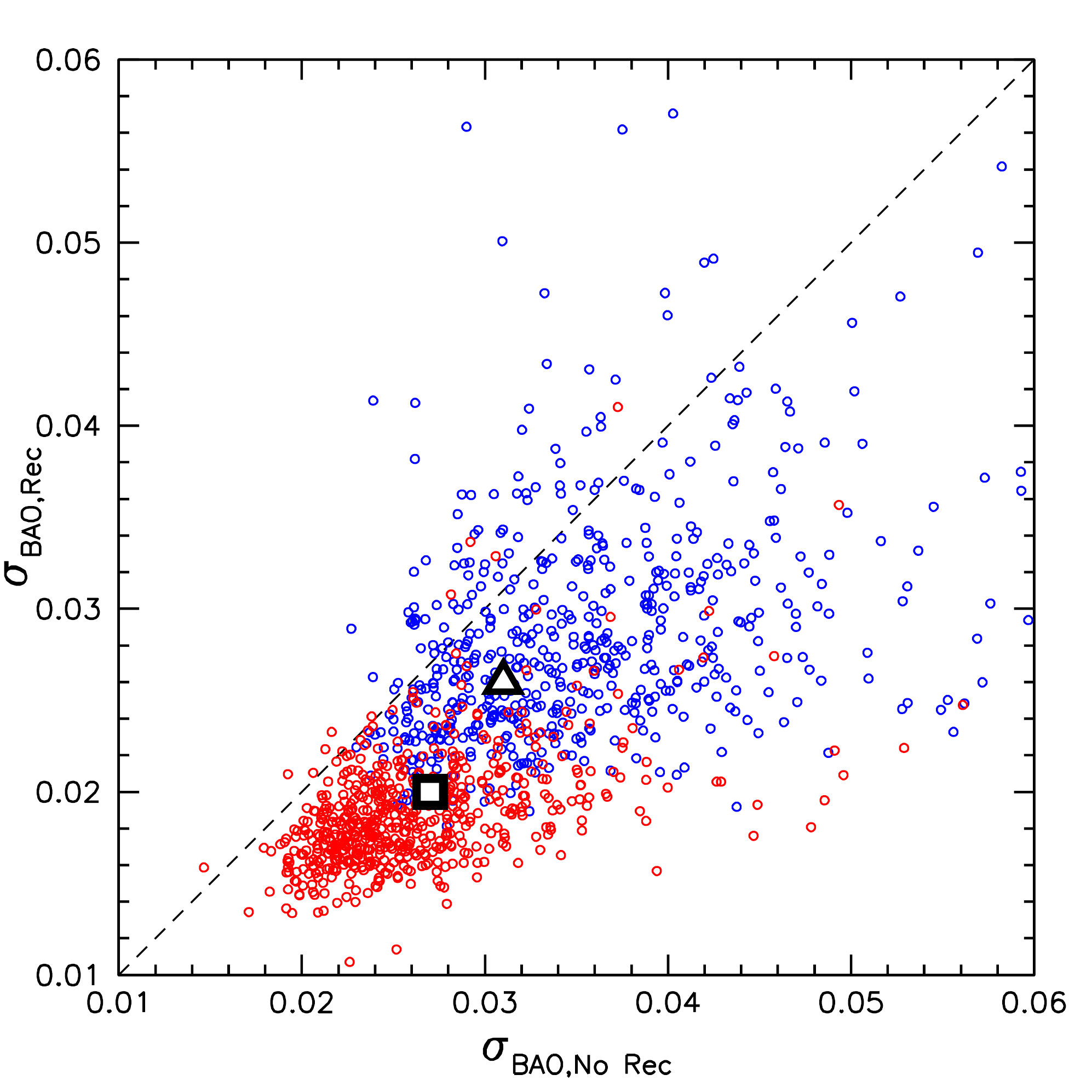}
  \caption{The uncertainty on the BAO position recovered from $\xi_0$ measurements after applying reconstruction (``Rec") versus those obtained before (``No Rec''). Points display the results from the 600 mock realizations of the Red (red points) and Blue (blue points) galaxy samples. The large black square and triangle represent the results for the Red and Blue CMASS data samples. Each result recovered from the CMASS data is within the locus of the uncertainties recovered from the mock realizations.
 }
  \label{fig:baosigreccom}
\end{figure}

\begin{figure}
\includegraphics[width=84mm]{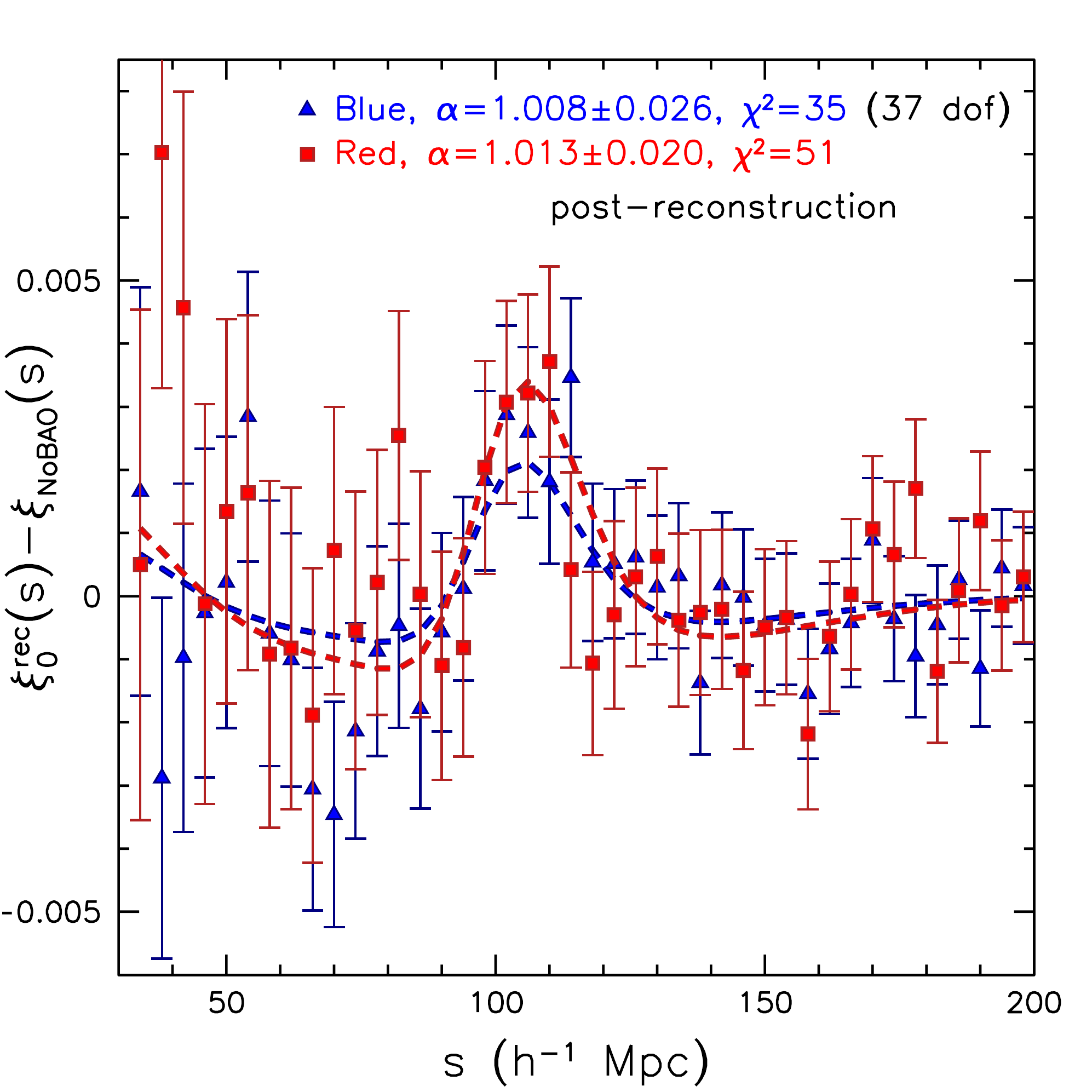}
  \caption{The measured $\xi^{\rm Rec}_0$ (points with error-bars) after applying reconstruction and the best-fit BAO model (dashed curves) for the Red (red) and Blue (blue) CMASS data samples. Each has had the smooth component of the best-fit model subtracted. Reconstruction has sharpened the BAO feature for both samples. The positions of the BAO feature found from the Red and Blue samples agree.}
  \label{fig:xi0BAOfitsrec}
\end{figure}

We apply reconstruction (see Section \ref{sec:rec}) to the Red and Blue samples (for both the data and the 600 mock realizations) and re-measure the BAO scale using $\xi^{\rm Rec}_0(s)$. Fig. \ref{fig:baosigreccom} displays the 600 of recovered uncertainties after applying reconstruction versus the uncertainty recovered prior to reconstruction for both the Blue (blue points) and Red (red points) mock samples. For both samples, the vast majority of mock realizations show an improvement in precision of the BAO measurement. As can be seen in Table \ref{tab:mockbao}, the improvement due to reconstruction larger for the Red samples, as the mean uncertainty has decreased by 32 per cent for the Red samples and by 24 per cent for the Blue samples. 

The measured $\xi^{\rm Rec}_0(s)$ of the Red and Blue data samples are compared to the best-fit models, both with the smooth component of the best-fit subtracted, in Fig. \ref{fig:xi0BAOfitsrec}. The $\chi^2$ of the best-fit for the Red sample is unusually high (51 for 37 dof), but, as noted in Section \ref{sec:RMcom}, this result is due mainly to the data at $s > 150 h^{-1}$Mpc. Reconstruction reduces the uncertainties on the Red and Blue data BAO measurements by 35 per cent and 19 per cent, similar to the mean effect found from the mock realizations. In Fig. \ref{fig:baosigreccom}, the data results are displayed using a black triangle for the Blue sample and a black square for the Red sample. Each are within the locus of points displaying the results recovered from the mock realizations. 

After applying reconstruction, both measurements of $\alpha$ have shifted only slightly from their pre-reconstruction values. The post-reconstruction Red and Blue data BAO measurements are clearly consistent, as they differ by only 0.005. We narrow the fit range to $50 < s < 160h^{-1}$Mpc and re-measure the BAO scale, denoting it $\alpha^{\prime}$. We find $\alpha^{\prime}_{\rm Red}=1.008\pm0.021$ and $\alpha^{\prime}_{\rm Blue}=1.002\pm0.025$. Each $\alpha^{\prime}$ measurement has shifted by $0.3\sigma$ compared to the fiducial $\alpha$ measurement. While coherent, such a shift alters none of our conclusions.

In summary, we find consistent BAO scale measurements for the clustering of the Red and Blue CMASS data samples and their cross-correlation, determined from both $P(k)$ and $\xi(s)$. The pair of measurements that disagree the most is $\alpha_{P,{\rm Red}} = 0.992 \pm 0.025, \alpha_{\xi,X} = 1.024\pm0.024$ and we find that 118 of the mock pairs have a larger $d_{\alpha}$ value. We find no observational evidence that measurements of the BAO position systematically depend on the properties of the galaxies one uses for the clustering measurement.

\section{RSD measurements}
\label{sec:RSD}
\subsection{Consistency}

\begin{figure}
\includegraphics[width=84mm]{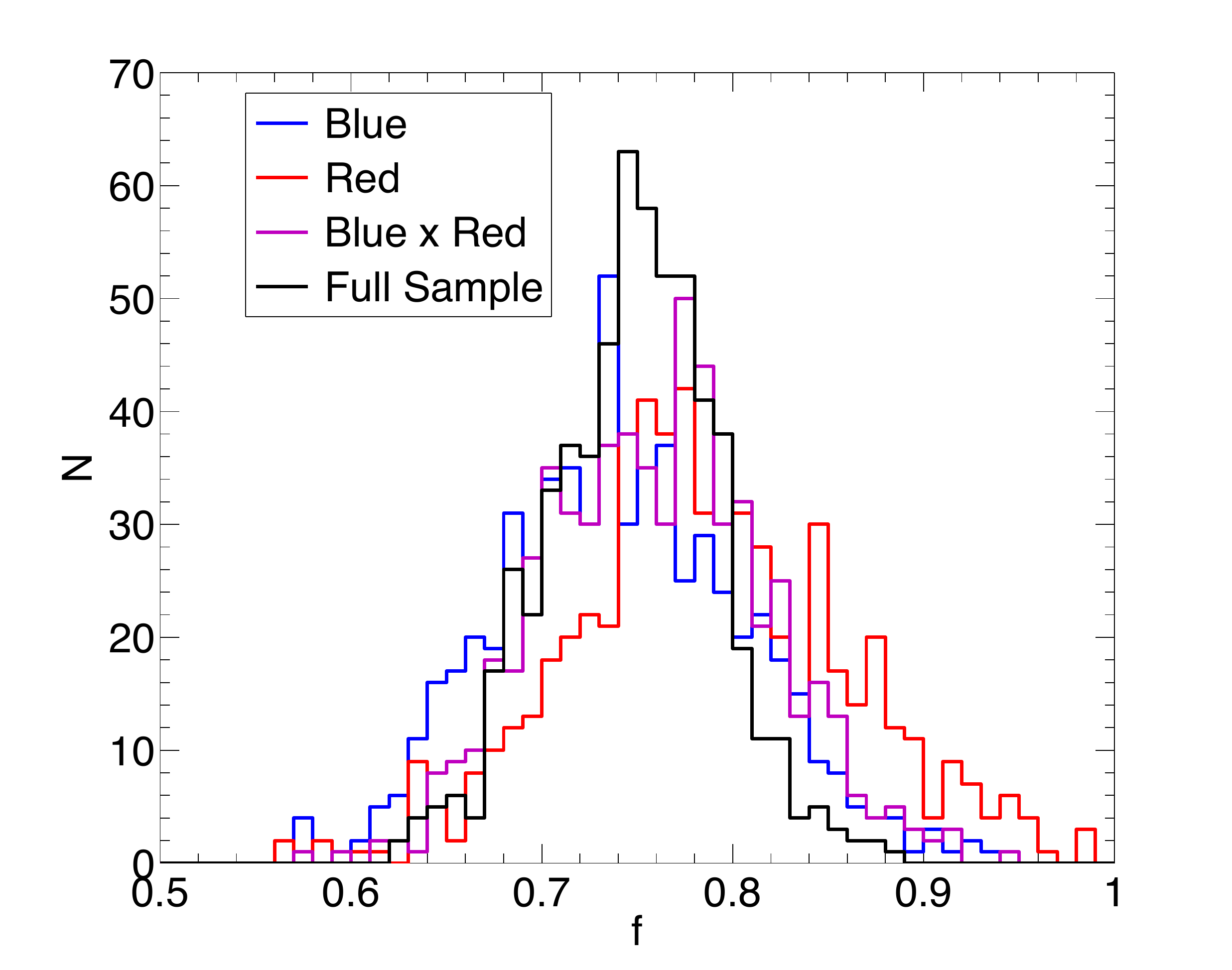}
  \caption{Histograms of $f$ values recovered from the $\xi_{0,2}$ of 600 mock realizations of the Blue (blue) and Red (red) samples and the cross $\xi_{0,2}$ (purple), as well as for the full CMASS sample (black). One can see that the Red and Blue distributions are slightly offset (by $f=0.018$) from each other.}
  \label{fig:rbc1}
\end{figure}

We test our method for fitting $f\sigma_8$, as described in Section \ref{sec:SGmod}, by applying it to individual mocks
to determine the best-fit values of bias, $b$, and the growth rate, $f$. All of our measurements
are based on fits to the $\xi_{0,2}$ measurements in the range $30 < s < 150
h^{-1}$Mpc. Since the modelling of AP distortions is more robust compared to
the RSD modelling, for simplicity we fix the fiducial cosmological model to the
input model of the mocks, and thus find $f$ for fixed $\sigma_8$. Fig.  \ref{fig:rbc1} displays histograms of the
distribution of best-fit growth rate measurements recovered from 600 mock realizations of the Blue and Red mock samples and their
cross-correlation. The distribution recovered from the Blue measurements is displayed in blue, that from the Red measurements in red, and that from the Red$\times$Blue measurements in purple. We also display the results using measurements recovered from realizations of the full CMASS sample in black. Note that the full CMASS sample contains more than twice as many galaxies as the sum of
our Red and Blue samples. For the mean and standard deviation of these distributions, we find $f_{\rm Blue} = 0.736 \pm 0.065$, $f_{\rm Red}
= 0.780 \pm 0.073$ and $f_{\rm cross} = 0.754 \pm 0.058$. These fits to $\xi_{0,2}$ data are summarized in Table \ref{tab:RSD}.

When we fit to the mean $\xi_{0,2}$ of all 600 mocks we find the best-fit values of $f_{\rm Blue}
= 0.724$ ($\chi^2 = 47/59 ~{\rm dof}$), $f_{\rm Red} = 0.776$ ($\chi^2 = 54/59~
{\rm dof}$ and $f_{\rm cross} = 0.752$ ($\chi^2 = 69/59 ~{\rm dof}$). These
values are consistent with the mean of the fits to individual mock samples and are biased by 2.7, 4.3 and 1.1 per cent with respect to the true input
value of the mocks. These results (at least qualitatively) appear consistent
with the findings of \cite{RW11}, where it was found that the model (the same as
we apply) over-predicts the value of $\xi_2$ for their low mass sample and
under-predicts the value of $\xi_2$ for their high mass sample, each by
$\sim$4\% at $s = 35h^{-1}$Mpc. The difference in $b$ values of the samples used
by \cite{RW11} is more extreme ($b_{\rm high} =2.8$, $b_{\rm low} = 1.4$ compared to our
$b_{\rm Blue} = 1.65\pm0.07$, $b_{\rm Red}=2.3\pm0.09$).

\begin{figure}
\includegraphics[width=84mm]{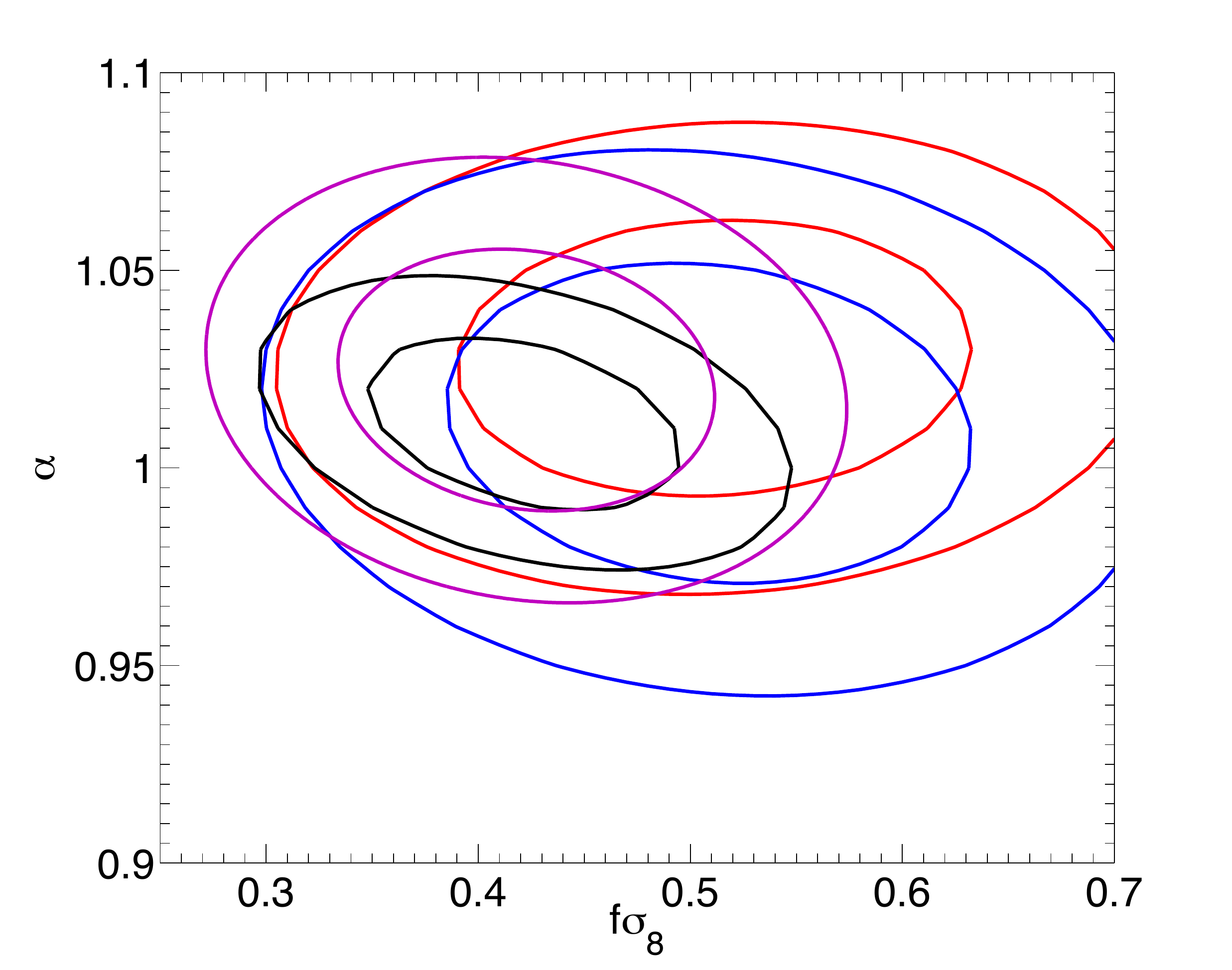}
  \caption{one and two $\sigma$ confidence level contours on $\alpha$ and $f\sigma_8$. The red and blue contours correspond to the Red and Blue samples, the purple curves are for their cross-correlation, and the black curves are for the full CMASS sample. Broadly, all samples yield consistent results.}
  \label{fig:alphfsig}
\end{figure}

Next we perform full fits to the Blue and Red data samples, now with AP parameters
kept free. We have derived quantities $\alpha$, $\epsilon$, and $f\sigma_8$
from full fits to the Red and Blue $\xi_{0,2}$ measurements. Fig.
\ref{fig:alphfsig} displays the 1 and 2$\sigma$ contours for the allowed
$\alpha$ and $f\sigma_8$ for the Red, Blue and full samples and the cross-correlation between the Red and Blue samples. The measurements $\alpha_{\rm Blue} = 1.011 \pm 0.028$, $\alpha_{\rm Red} = 1.028 \pm
0.024$ and $\alpha_{\rm Cross} = 1.022 \pm 0.023$ are consistent with each other
and those we find when fitting only the BAO feature (see Table
\ref{tab:mockbao}), but the full shape information has allowed a small
improvement in the uncertainty on the BAO-only fit. 

\begin{table*}
\begin{minipage}{7in}
\caption{The statistics of growth and distance parameters recovered from the mock and data Red and Blue anisoptropic clustering measurements. The parameter $\langle f \rangle$ is the mean $f$ value determined from 600 mock realizations of each sample, $S_{f} = \sqrt{\langle(f-\langle f\rangle)^2\rangle}$ is the standard deviation of the best-fit $f$ values, $\langle \chi^2 \rangle$ is the mean minimum $\chi^2$ value, and ``CMASS $f\sigma_8$, $\alpha$, $\epsilon$'' is the full set of measurements for the data samples. The ``Combined'' data are recovered from the combination of all Red, Blue, and cross-correlation pair counts. The ``Opt. Combined'' data are the optimal combination of $f$ measurements, determined using the covariance between the recovered $f$ of each sample determined using the mock samples. The ``Full'' data is the full CMASS sample, which contains more than twice as many galaxies as the Red and Blue samples combined.}
\begin{tabular}{lcccl}
\hline
\hline
Case  &  $\langle f \rangle$ & $S_{f}$ & $\langle \chi^2 \rangle$/dof & CMASS $f\sigma_8$, $\alpha$, $\epsilon$, $\chi^2$/dof \\
\hline
Red & 0.780 & 0.073  & 61/61 & 0.509$\pm$0.085, 1.028$\pm$0.024, -0.032$\pm$0.024, 48/54 \\
Blue & 0.736 & 0.065 & 61/61 & 0.511$\pm$0.083, 1.011$\pm$0.028, -0.034$\pm$0.031, 70/54 \\
Cross & 0.754 & 0.058 & 62/61 & 0.423$\pm$0.061, 1.022$\pm$0.023, -0.023$\pm$0.024, 37/54 \\
Combined & 0.751 & 0.056 & 63/61 & 0.464$\pm$0.059,1.020$\pm$0.022, -0.029$\pm$0.023 50/54 \\
Opt. Combined & 0.755 & 0.053 & - & 0.443$\pm$0.055, -, - ,- \\
Full & 0.743 & 0.0440 & 61/61 & 0.422$\pm$0.052, 1.011$\pm$0.015, 0.002$\pm$0.018, 60/54 \\
\hline
\label{tab:RSD}
\end{tabular}
\end{minipage}
\end{table*}

\begin{figure}
\includegraphics[width=84mm]{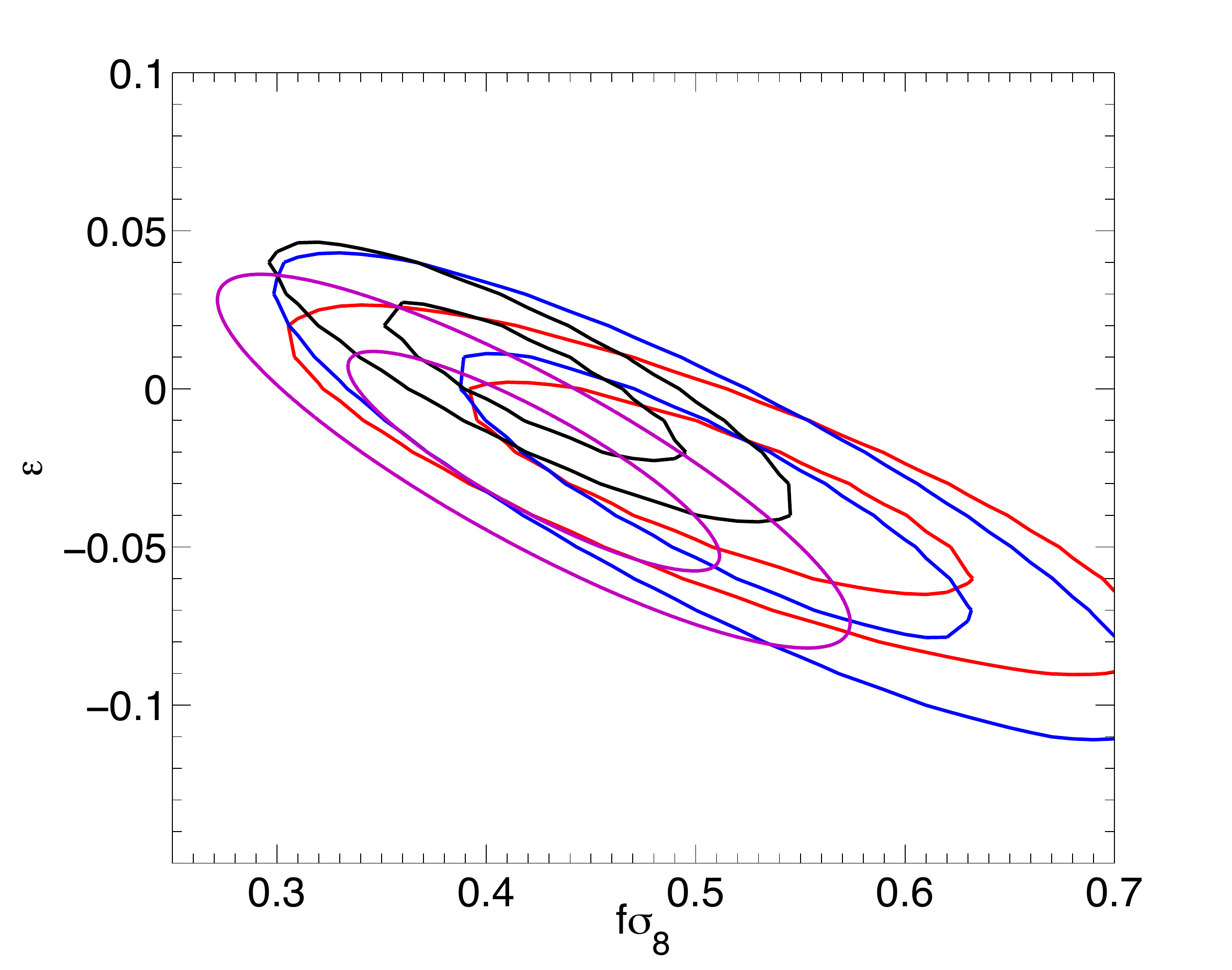}
  \caption{Same as Fig \ref{fig:alphfsig}, but for $\epsilon$ (defined in Eq. \ref{eq:ep}) and $f\sigma_8$. The Red and Blue samples yield consistent results. A slight tension (1.4$\sigma$) is observed between the $\epsilon$ value of the Red sample and that of the full sample.}
  \label{fig:epfsig}
\end{figure}

Our fitting procedure yields $f\sigma_{8,\rm Blue} = 0.509 \pm 0.085$, $f\sigma_{8,\rm Red} = 0.511 \pm
0.083$ and $f\sigma_{8,\rm Cross} = 0.423 \pm 0.061$ for the data samples. We see no evidence of the 7\% difference in growth values found in the mock samples, however the difference could easily be hidden in the noise, given we achieve 17\% precision. The results obtained from
the Blue and Red samples are somewhat higher than the results obtained from the
fits to the full sample ($f\sigma_{8,\rm Full} = 0.422 \pm 0.051$). The Red and Blue samples are each less than one quarter the size of the full sample. Thus the differences between the Red and Blue $f\sigma_8$ values and that obtained from the full sample are of order of 1$\sigma$ and therefore not statistically significant. The results obtained from the cross-correlation of the two samples are in a good
agreement with the full sample. Factoring in the covariance found between the cross-correlation measurements and the Red/Blue measurements of our mock samples, the differences between the Red and Blue $f\sigma_8$ and the cross-correlation result both represent a 1.3$\sigma$ discrepancy.

Fig. \ref{fig:epfsig} displays the 1 and 2$\sigma$ contours for the allowed
$\epsilon$ and $f\sigma_8$ for the Red and Blue samples. We find
$\epsilon_{\rm Red} = -0.032 \pm 0.024$, $\epsilon_{\rm Blue} = -0.034 \pm 0.031$ and
$\epsilon_{\rm Cross} = -0.023 \pm 0.024$,
fully consistent with each other. The value recovered from the full sample ($\epsilon = 0.002 \pm 0.018$) is
within 1.43 $\sigma$ of the Red sample and 1.16 $\sigma$ of the Blue sample.

Overall, the triplet of measured values of $f\sigma_8$, $\alpha$ and $\epsilon$
is consistent between Red and Blue samples and their cross-correlation and that of the full sample. For each pair of triplets, there is a 3$\times$3 covariance matrix for the data vector $d = [f\sigma_8,\alpha,\epsilon]_1-[f\sigma_8,\alpha,\epsilon]_2$. We find the $\chi^2$ testing $d$ for each pair of triplets against the model $d_m =  [0,0,0]$.
For the Red and Blue samples we find $\chi^2 = 0.2$, between Blue and
full samples we find $\chi^2 = 1.1$ and between Red and full samples we find $\chi^2 = 2.1$. Given three degrees
of freedom, all are consistent within 1$\sigma$.

\subsection{Combining Tracers}
\label{sec:2t}

\cite{McDSel09} have demonstrated that if two or more tracers with significantly
different bias trace the same underlying distribution of matter it is possible to
significantly strengthen derived measurements of cosmic growth rate by the
virtue of the fact that the samples share the cosmic variance contribution to
the errors. To study the applicability of this method to our CMASS sample and
the expected improvement in the measurements, we extract the growth measurements
from 600 mocks of Blue and Red samples and examine the distribution of best-fit
values.

\begin{figure}
\includegraphics[width=84mm]{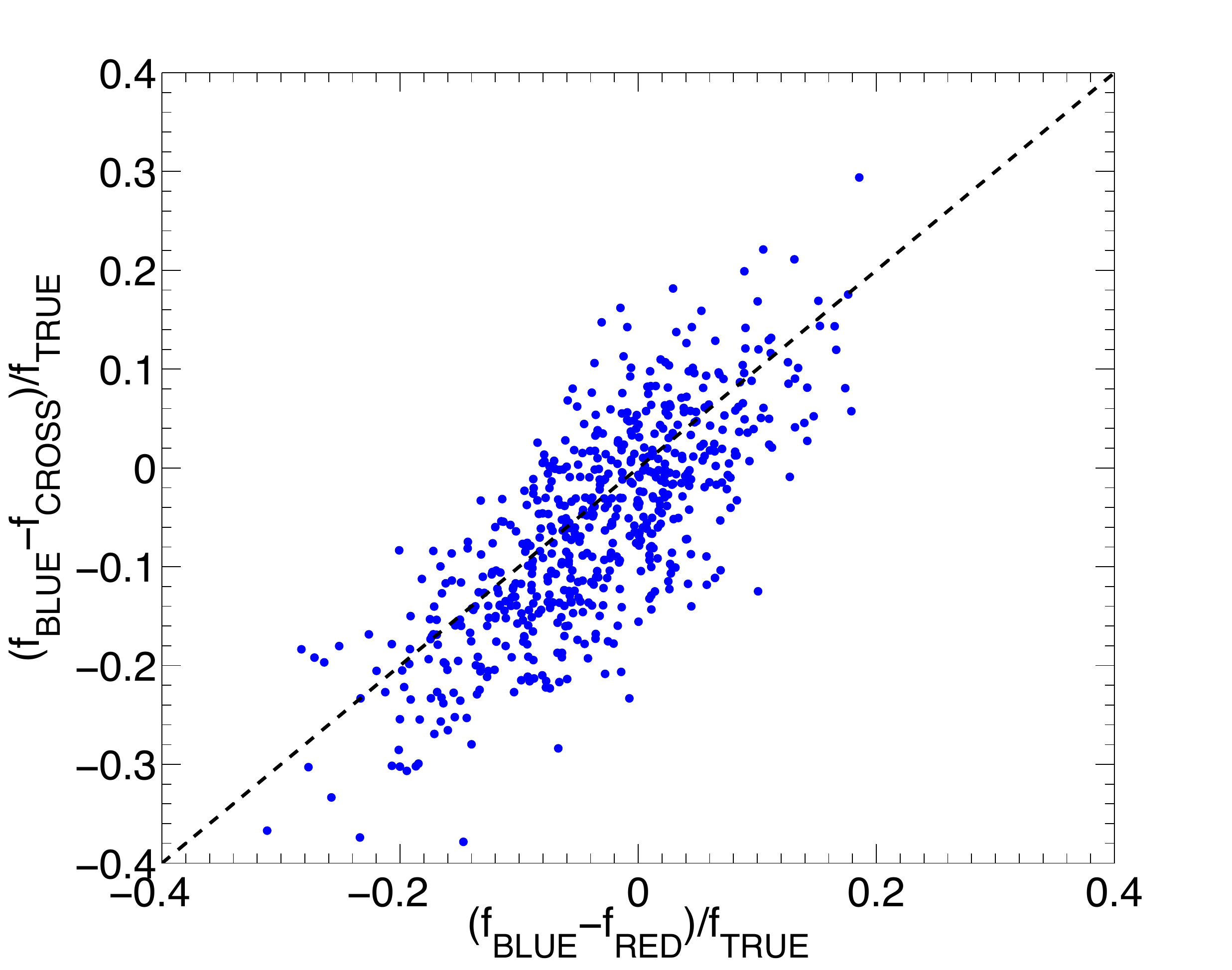}
  \caption{The offset between $f$ values derived from 600 mock realizations of Blue and Red samples and
  the cross-correlation of each realization. A strong correlation between the recovered $f$ values is observed.}
  \label{fig:rbdisp}
\end{figure}

\begin{figure}
\includegraphics[width=84mm]{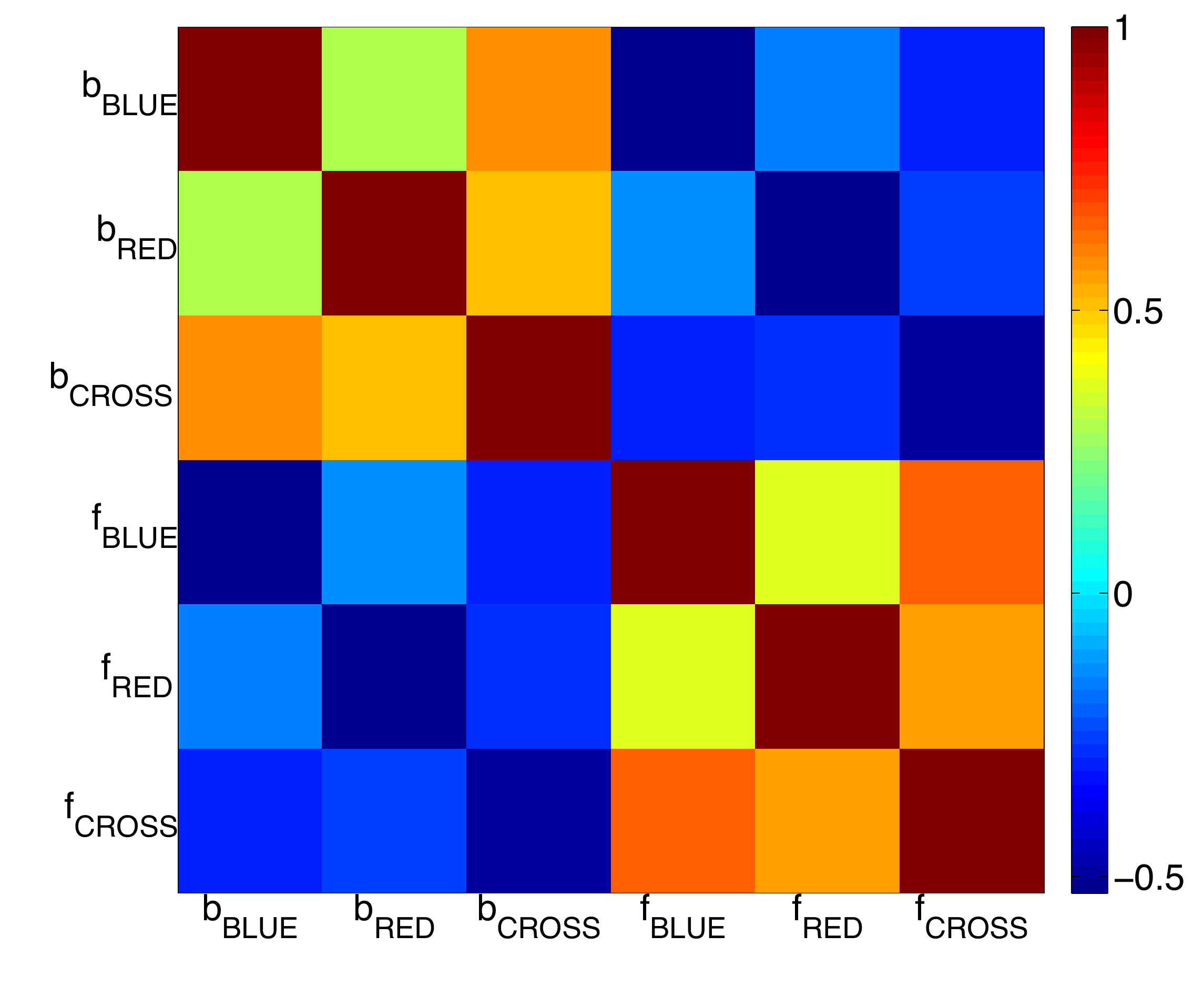}
  \caption{The reduced covariance matrix of the bias, $b$, and growth parameter, $f$, for the Red and Blue samples and their cross-correlations, as determined from fitting the 600 mock samples. We use this covariance matrix to optimally combine our results from the Red and Blue samples and their cross-correlation to produce an optimized $f\sigma_8$ measurement.}
  \label{fig:bfcov}
\end{figure}

Fig.~\ref{fig:rbdisp} displays the offset between the $f$ values obtained from
the Blue and Red samples extracted from the same underlying dark matter
distribution and their cross-correlation. The values extracted from the individual mocks can be offset by as much as 40
per cent, but the measurements are strongly correlated on average. For the Blue,
Red samples and their cross-correlation we obtain measurements of $b$ and $f$ for each realization and construct
their $6\times6$ covariance matrix. This (reduced) covariance matrix is shown
in Fig \ref{fig:bfcov}. The correlation between the Red and Blue $f$ measurements is 0.37.

To take advantage of the fact that the estimates of growth are correlated by
the virtue of having almost identical cosmic variance and the fact that the
bias of the cross sample must satisfy $b_{\rm cross} = \sqrt{b_{\rm Red}b_{\rm
Blue}}$, we fit to these six measurements ($f_{\rm Blue}, b_{\rm Blue}$, $f_{\rm Red}, b_{\rm Red}$,
$f_{\rm Cross}, b_{\rm Cross}$) a three-parameter model $p = (f, b_{\rm Blue}, b_{\rm Red})$.
Applied to the distribution of mock $b,f$ values, this fit produces the best-fit value and one $\sigma$ standard deviation of
$f = 0.755 \pm 0.053$. When instead the constraints are derived from the $\xi_{0,2}$ calculated from the sum of Red, Blue, and cross pair counts, we find a mean and standard deviation of $f=0.751\pm$0.056. Thus, by splitting the data sample and re-weighting the results we obtain a modest six per cent improvement in the recovered value of $f$.

 In our particular case, the gain in the estimate of $f$ is small mainly because the errors of individual
measurements are dominated by the shot-noise at small scales. Given greater number densities (this is not possible with the BOSS galaxy sample while maintaining the same difference in bias) the correlation between the Red and Blue samples would be larger and thus a greater gain in the precision of $f$ would be achievable. The value improved the most by the combination of samples is the ratio of biases of two samples. Using only the Blue and Red samples (but accounting for their covariance) $b_{\rm Red}/b_{\rm Blue} = 1.39 \pm 0.05$. Using the full optimized data set, we recover $b_{\rm Red}/b_{\rm Blue} = 1.39 \pm 0.04$, a 20 per cent improvement.

The most consistent approach to extracting the growth rate constraints from the data
would be to fit to all three measured correlation functions simultaneously.
This, however, would require accurately estimating covariance matrices of rank
of order of few hundred. Even with 600 mocks this exercise would induce large error on
our final results (see, e.g., \citealt{PerCov}). Instead, we assume that the
three individual measurements of $b$ and $f\sigma_8$ from the Blue and Red samples and
their cross-correlation are not biased and adopt the 6x6 reduced covariance matrix
computed from the mocks. This yields our optimized measurement, from the Red and Blue data samples, of $f\sigma_{8,\rm comb} = 0.443 \pm 0.055$. 

Although our optimized results are not as good as what is obtained for the full
CMASS sample ($f\sigma_{8,\rm Full} = 0.422 \pm 0.051$) one should keep in mind that, combined, our Blue and Red samples contain less
than half of all the galaxies in the full sample, and the uncertainty on our result is
less than 10 per cent greater than what is achieved with the full sample. This implies that one could
obtain the best CMASS $f\sigma_8$ measurements by using all of the CMASS data
and finding the optimal way in which to split into separate samples. To obtain
robust results to be used for a precise test of General Relativity a more
accurate modelling of redshift-space distortions for low and high bias samples
is required, such as presented in the recent results of \cite{WRW13}.

\section{Conclusions}
\label{sec:con}

We find no detectable difference in distance scale or growth measurements obtained from DR10 BOSS CMASS galaxies when the sample is split by colour. This result is in agreement with theoretical predictions (e.g., \citealt{Pad09,RW11}) and the results we obtain from mock samples. These measurements provide additional evidence that BAO and RSD measurements are precise and robust probes of Dark Energy.

We have selected two subsets of BOSS CMASS galaxies based on their $k+e$ corrected $i$-band absolute magnitudes and $[r-i]_{0.55}$ colours. Our selection yields a Blue sample with the 23 per cent bluest galaxies and a Red sample containing the 32 per cent most luminous of the galaxies not in the Blue sample. The samples have similar $n(z)$ (see Fig. \ref{fig:nz}) and have a factor of two difference in clustering amplitude. 

We have created 600 mock realizations of each of our samples by sub-sampling each full CMASS mock realization based on halo mass in order to reproduce the observed clustering. In Section \ref{sec:RMcom}, we show that the clustering of the mock samples is a good match to the observed clustering. Fixing the background cosmology to our fiducial one and fitting for a constant, linear (real-space) bias in the range $20 < s < 200$, we find $b=2.30\pm0.09$ for the Red sample and $b=1.65\pm0.07$ for the Blue. For the cross-correlation, we find $b=1.96\pm0.05$, close to the geometric mean of the two best-fit values, as expected if a simple linear bias model is appropriate.

We have measured the BAO scale (parameterized by $\alpha$) from the clustering of each of the mock realizations. We find that the mean $\alpha$ recovered from Red mock samples is 0.003 larger than that of the Blue mock samples. This difference is consistent with the bias-dependent shift found in \cite{Pad09}. After applying reconstruction to each sample, the difference is reduced to less than a 1$\sigma$ discrepancy.

We have measured the amplitude of the velocity field ($f$) from $\xi_{0,2}$ measurements of each mock realization. The $f$ values recovered from the Blue sample are biased to low $f$ values by 2.7 per cent and that the values of the Red sample are biased to high $f$ values by 4.3 per cent. Our results are based on the model of \cite{RW11} and the bias is consistent with their findings.

The expected difference between the BAO scale measured from the Blue sample and that measured from the Red sample is less than 10 per cent of the standard deviation of the difference between Red and Blue BAO measurements and we therefore do not expect to be able to detect any difference in the CMASS data samples. Indeed, the BAO measurements of the Red and Blue CMASS samples are statistically indistinguishable (see Table \ref{tab:mockbao}).

The expected difference between the $f\sigma_8$ values recovered from the Blue and Red samples is 33 per cent of the standard deviation of the difference found from the mock realizations. While larger than the discrepancy expected for the BAO measurements, we still do not expect to find any statistically significant tension. Indeed, the results are perfectly consistent ($f\sigma_{8,Blue} = 0.509 \pm 0.085$; $f\sigma_{8,Red} = 0.511 \pm0.083$; and they would remain consistent if a correction factor for the bias was applied). For the final CMASS data set (which will be roughly twice as large), the expected discrepancy would increase to 50 per cent, but we expect usage of the refined model of \cite{WRW13} will significantly reduce the bias.

We have used the covariance between mock measurements of $f,b$ values obtained from the Red, Blue, and their cross-correlation $\xi_{0,2}$ in order to obtain the optimal combination to produce a single $f\sigma_8$ measurement. Applying this to the data, we find $f\sigma_8 = 0.443\pm0.055$. This result compares well to what is achieved from the full CMASS sample ($f\sigma_8 = 0.422\pm0.051$) despite the fact we have used less than half of the total sample to obtain our result. These results suggest that producing the optimal measurement of $f\sigma_8$ using BOSS CMASS galaxies can be accomplished by combining measurements of the Red and Blue samples used herein as well as the remaining 53 per cent of CMASS galaxies we have omitted from this analysis (and all of their cross-correlations).

The Red and Blue sample we have defined may be used for further tests. Modelling the effect of massive neutrinos on the measured power spectrum is somewhat degenerate with the non-linear bias model one uses (see, e.g., \citealt{SwansonNeut,Zhao12neut}), and the robustness of the modelling that is applied can be tested by using different galaxy populations (as done in \citealt{SwansonNeut} for galaxies from the SDSS main sample). Similar tests can be applied to the same BOSS galaxy samples used herein. The large-scale $P(k)$ measurements of our samples can also be combined to produce more robust measurements of local non-Gaussianity, given that the signal is expected to be proportional to the bias of the sample. Further, given this measurement relies on the largest scales, the covariance between the Red and Blue samples will be higher and thus allow greater gain from two-tracer method. In future analyses, splitting samples by colour may simultaneously test robustness and increase precision.

\section*{Acknowledgements}
We thank the anonymous referee for comments that helped improve the paper. AJR is thankful for support from University of Portsmouth Research Infrastructure Funding. LS is grateful to the European Research Council for funding. AB is grateful for funding from the United Kingdom Science \& Technology Facilities Council (UK STFC). WJP acknowledges support from the UK STFC through the consolidated grant ST/K0090X/1, and from the European Research Council through the ÒStarting Independent ResearchÓ grant 202686, MDEPUGS.

Mock catalog generation, correlation function and power spectrum calculations, and fitting made use of the facilities and staff of the UK Sciama High Performance Computing cluster supported by the ICG, SEPNet and the University of Portsmouth.

Funding for SDSS-III has been provided by the Alfred P. Sloan
Foundation, the Participating Institutions, the National Science
Foundation, and the U.S. Department of Energy Office of Science.
The SDSS-III web site is http://www.sdss3.org/.

SDSS-III is managed by the Astrophysical Research Consortium for the
Participating Institutions of the SDSS-III Collaboration including the
University of Arizona,
the Brazilian Participation Group,
Brookhaven National Laboratory,
Cambridge University ,
Carnegie Mellon University,
Case Western University,
University of Florida,
Fermilab,
the French Participation Group,
the German Participation Group,
Harvard University,
UC Irvine,
Instituto de Astrofisica de Andalucia,
Instituto de Astrofisica de Canarias,
Institucio Catalana de Recerca y Estudis Avancat, Barcelona,
Instituto de Fisica Corpuscular,
the Michigan State/Notre Dame/JINA Participation Group,
Johns Hopkins University,
Korean Institute for Advanced Study,
Lawrence Berkeley National Laboratory,
Max Planck Institute for Astrophysics,
Max Planck Institute for Extraterrestrial Physics,
New Mexico State University,
New York University,
Ohio State University,
Pennsylvania State University,
University of Pittsburgh,
University of Portsmouth,
Princeton University,
UC Santa Cruz,
the Spanish Participation Group,
Texas Christian University,
Trieste Astrophysical Observatory
University of Tokyo/IPMU,
University of Utah,
Vanderbilt University,
University of Virginia,
University of Washington,
University of Wisconson
and Yale University.

\label{lastpage}

\end{document}